\documentclass[aps, prfluids, reprint, onecolumn]{revtex4-2}
\usepackage{amsfonts} 

\usepackage{stmaryrd}
\usepackage{amsmath}
\usepackage{amssymb}
\usepackage{float}
\usepackage{blindtext}
\usepackage{lineno}
\usepackage{mydef}
\usepackage{cancel}
\usepackage{bm}
\usepackage[usenames, dvipsnames]{color} 
\usepackage{dcolumn}
\usepackage{epic} 
\usepackage{epsfig}
\usepackage{esdiff}
\usepackage{graphicx}
\usepackage[breaklinks,colorlinks = true,linkcolor = blue,urlcolor=blue,citecolor=red]{hyperref}
\usepackage[caption=false]{subfig}
\begin{document}
\title{Active Sculpting: morphodynamics of internally stirred active vesicles}
\author{Devadevan M M}
\email{devadevan.mm@icts.res.in}
\affiliation{International Centre for Theoretical Sciences, Tata Institute of Fundamental Research, Bengaluru 560089, India}
\author{Brato Chakrabarti}
\email{brato.chakrabarti@icts.res.in}
\affiliation{International Centre for Theoretical Sciences, Tata Institute of Fundamental Research, Bengaluru 560089, India}

\begin{abstract}
Active flows are often confined by soft, deformable boundaries that they can push against, reshape, and in turn get reorganized by. Motivated by this interplay, we combine theory and numerical simulations to analyze the morphodynamics of an elastic vesicle enclosing a dilute suspension of extensile active particles. We model the two-dimensional vesicle as a weakly extensible elastic ring with bending rigidity, immersed in a Stokesian fluid, and describe the confined active suspension using a kinetic theory. The vesicle and the suspension are coupled hydrodynamically, through the kinematic and dynamic boundary conditions at the interface, and energetically, through a single alignment interaction that simultaneously aligns active particles with the membrane and the membrane with the particles. We show that this interaction reduces to a weak-anchoring boundary condition familiar from passive liquid crystals and derive a closed-form quiescent equilibrium state in which the nematic order in the suspension organizes into a vortex with a $+1$ defect core. A stability analysis of this state predicts an ordered hierarchy of activity thresholds at which successive shape modes of the vesicle are destabilized, as well as a purely passive, anchoring-driven instability of floppy vesicles. Nonlinear simulations, performed with a volume-conserving immersed boundary method, reveal that each threshold crossing gives rise to a distinct emergent state: a static spindle with boundary-pinned defects, a rotor that tank treads against its own rotation, a fluctuating spinner, and a turbulent state in which the vesicle acts as a self-stirring compartment. Beyond demonstrating how boundary deformability sculpts, and is sculpted by, confined active matter, our results open the door to study hydrodynamic interactions between such active vesicles and to use membrane fluctuations as a probe of activity of suspensions.
\end{abstract}

\maketitle

\section{Introduction}\label{sec:intro}

Active fluids such as suspensions of swimming bacteria or reconstituted mixtures of cytoskeletal filaments and molecular motors spontaneously generate large-scale flows powered by their microscopic constituents~\cite{marchetti2013,doostmohammadi2018}. In biological settings, these flows rarely act in isolation: they develop within, and push against, soft deformable interfaces, and this interaction is central to function across scales. Examples include flows and stresses generated by the actomyosin cortex that mold the cell surface and constrict it along the cytokinetic furrow~\cite{mayer2010, turlier2014, rajput2025active}; cytoplasmic flows driven by cytoskeleton assemblies organize organelles and subcellular compartments in an evolving cell interior~\cite{nazockdast2017cytoplasmic}; and inside the nucleus, the activity of transcribing DNA drives coherent, micron-scale motions of the chromatin fiber~\cite{zidovska2013, saintillan2018extensile, weady2024, rautu2026}. Synthetic and reconstituted systems display the same interplay: interfaces separating an active from a passive fluid exhibit giant fluctuations and propagating waves~\cite{adkins2022dynamics, gulati2024traveling, sciortino2025active}, rafts of active colloidal particles can propel droplets \cite{kato2026active}, and flexible polymers suspended in an active bath can stiffen, buckle, oscillate, or be continually deformed and transported by the dynamics around them~\cite{kikuchi2009buckling, weady2024, valei2025scaling}.

Soft enclosures realize this bidirectional coupling between active flows and deformation in its most consequential form: the active material shapes the enclosure that confines it, while the boundary constrains the activity within. The canonical example of this setting is active droplets that spontaneously break symmetry to swim and crawl~\cite{callanjones2008, hawkins2011spontaneous, tjhung2012, gao2017self, young2021, nejad2023spontaneous, sessa2026interfacial} and to elongate and divide~\cite{giomi2014,weirich2019}; similarly, vesicles and drops coated with an active nematic film deform in concert with the motion of their surface defects~\cite{keber2014, metselaar2019,  firouznia2025self}. Encapsulation experiments now probe this interplay directly: cytoskeletal networks reconstituted inside giant unilamellar vesicles remodel and bleb their host~\cite{loiseau2016}, actomyosin rings constrict vesicles along a well-defined furrow~\cite{litschel2021}, and active colloids or bacteria enclosed in floppy vesicles drive large persistent deformations~\cite{takatori2020active, vutukuri2020, le2022encapsulated, kato2026active}. Motivated by these systems, this paper addresses a minimal version of the enclosure problem: how a confined suspension of momentum-conserving active particles sculpts the soft elastic boundary that contains it, and how the evolving shape, in turn, reorganizes the activity within.

Theoretical efforts to understand this interplay have so far focused largely on \emph{dry} settings, in which self-propelled particles interact with a deformable enclosure through direct collisions with the hydrodynamics of the suspending solvent neglected. Such models of active Brownian particles confined in deformable vesicles and rings capture shape and displacement fluctuations of the enclosure, spontaneous motility, and the tether-like protrusions observed in  some experiments~\cite{paoluzzi2016, vutukuri2020, peterson2021vesicle}. Many of the biological and reconstituted systems described above are, however, fundamentally \emph{wet}: stresses exerted by the active particles are carried by the suspending fluid, which in turn drives long-ranged flows, and the enclosed suspension communicates with its boundary through these hydrodynamic interactions rather than solely by contact. Rigid confinement is already well known to dictate the nature of such emergent active flows: bacterial suspensions self-organize into vortex lattices in pillar arrays~\cite{wioland2016ferromagnetic}; microtubule-based active fluids confined in channels and racetracks spontaneously pump and circulate coherently~\cite{doostmohammadi2017onset, wu2017,theillard2017geometric}; and active nematics in rigid disks exhibit a sequence of defect-ordered states en route to confined turbulence~\cite{theillard2017geometric, shendruk2017,norton2018,opathalage2019,hardouin2019}.

What happens in the presence of deformable boundaries is a question we are only beginning to understand. Recent studies have revealed that passive droplets immersed in confined active nematics are linearly unstable to interfacial perturbations driven by active and elastic stresses~\cite{alert2022fingering, dhar2025}, though the emergent states beyond onset remain unknown; similarly, soft bodies suspended in active nematic baths also deform under the combined action of anchoring-induced elastic stresses and activity~\cite{chandler2024, chandler2025}. However, how an enclosed active fluid and the elasticity of the boundary that confines it conspire, beyond the onset of instability, to select morphology and govern the topology of the interior flows still remains poorly understood. Here, we develop this picture for a minimal model of an internally stirred two-dimensional elastic vesicle, and show that the competition between internal activity and bending rigidity organizes a sequence of emergent morphodynamic states in which the vesicle is actively sculpted by the suspension it confines.

The paper is organized as follows. In Sec.~\ref{sec:model} we develop a minimal momentum-conserving model of an internally stirred vesicle: a closed, nearly inextensible elastic ring that resists bending and encloses a dilute extensile suspension described by a kinetic theory~\cite{saintillan2008prl}.  The ring exchanges momentum with the interior and exterior fluids as in classical fluid--structure interaction problems, while steric interactions between the suspension and the membrane give rise to a mutual alignment coupling at the interface. We then outline an accurate numerical method, built on the immersed boundary framework~\cite{peskin2002,peskinprintz1993}, that exactly preserves the volume of the confinement. In Sec.~\ref{sec:lsa} we show how our alignment interaction maps onto classical models of weak anchoring in liquid crystals, derive the quiescent base state of the confined suspension, and perform a linear stability analysis that predicts an ordered hierarchy of activity thresholds at which successive shape modes destabilize. Section~\ref{sec:nonlinear} follows these instabilities into the nonlinear regime, where we numerically identify a sequence of emergent morphodynamic states of increasing dynamical complexity, ranging from stationary deformed shapes to rotating, fluctuating, and ultimately turbulent vesicles. We conclude in Sec.~\ref{sec:conclusion} with a discussion of our results and an outlook on future directions.
\section{Modeling and simulation of 2D active vesicles}\label{sec:model}
We model 2D active vesicles as closed elastic rings with bending and stretching elasticity, suspended in an incompressible Stokesian fluid. The vesicle encloses an active suspension of rod-shaped particles that generate dipolar extensile stresses. We now describe each component of this active fluid--structure interaction (FSI) model, beginning with the hydrodynamics.

\subsection{Hydrodynamics}
At the micron scale of interest, fluid inertia is negligible, and the exterior of the vesicle is governed by the incompressible Stokes equations
\begin{equation}
    -\nabla q^\text{e} + \mu \Delta \bu^\text{e} = \mathbf{0}, \quad \nabla \cdot \bu^\text{e} = 0,
\end{equation}
where $q^\me(\bx)$ is the pressure, $\bu^\me(\bx)$ is the fluid velocity, $\mu$ is the viscosity; we have used the superscript $`\me'$ to denote quantities exterior to the vesicle. We define the stress tensor of the exterior fluid as $\bSigma^\me = -q^\me \bI + 2\mu\mathbf{E}^\me$, where $\mathbf{E}^\me$ is the symmetric rate-of-strain tensor, so the momentum balance reduces to $\nabla\cdot\bSigma^\me=\mathbf{0}$. The vesicle interior confines an active rod suspension. The leading-order hydrodynamic signature of force-free active particles that make up the suspension is a force dipole. A suspension of such active rods correspondingly gives rise to an active stress $\bSigma^\ma$, and the momentum balance can be written as
\begin{equation}\label{eq:actfl}
    -\nabla q^\mi + \mu \Delta \bu^\mi + \nabla \cdot \bSigma^\ma = \mathbf{0}, \ \ \nabla \cdot \bu^\mi = 0.
\end{equation}
Here, for simplicity, we assume the interior and exterior fluids to have identical viscosity; as before, the superscript $`\mi'$ denotes quantities interior to the vesicle. It is convenient to define a total stress tensor for the suspension as $\bSigma^\mT = \bSigma^\mi + \bSigma^\ma$, where $\bSigma^\mi$ encodes the passive Newtonian contribution. We now turn to our description for the active suspension.

The suspension contains $\mathcal{N}$ active particles confined within a two-dimensional domain $\Omega^\mi \subset \mathbb{R}^2$, where $\Omega^\mi$ denotes the interior of the vesicle. To describe the dynamics of the suspension, we use a Smoluchowski kinetic theory framework. In this approach, the statistical configuration of the suspension is described by a probability distribution function $\Psi(\bx, \bp, t)$, representing the probability of finding a particle at position $\bx \in \Omega^\mi$ with orientation $\bp \in \mathbb{S}^1$ such that $\int_{\Omega^\mi}\int_{\mathbb{S}^1} \Psi \md \bp \md \bx = \mathcal{N}$ \cite{saintillan2008prl, weady2024}. The evolution of the particle distribution is governed by a conservation equation
\begin{equation}
\label{eq:smoluchowski}
    \frac{\partial \Psi}{\partial t} + \nabla_\bx \cdot
    (\dot{\bx} \Psi) + \nabla_{\bp} \cdot (\dot{\bp} \Psi) = 0,
\end{equation}
where $\nabla_\bx$ is the spatial gradient, $\nabla_\bp = (\bI -\bp \bp) \cdot \partial_\bp$ is the gradient operator on the unit circle, and $\{\dot{\bx},\dot{\bp}\}$ are the translational and rotational fluxes describing the dynamics of individual active particles. For a dilute suspension of rod-like particles, these fluxes are given by
\begin{align}
    \dot{\bx} &= \bu^\mi(\bx) - d_T \nabla_\bx \ln \Psi,  \\
    \dot{\bp} &= (\bI - \bp\bp) \cdot \nabla \bu^\mi \cdot \bp
    - d_R \nabla_{\bp} \ln \Psi, \label{eq:jeff}
\end{align}
where $\bu^\mi(\bx)$ is the mean-field suspension velocity field interior to the vesicle and $\{d_T,d_R\}$ encodes the translational and rotational diffusivity of the particles. We highlight that in this work, we restrict ourselves to particles that are immotile but generate active dipolar stresses. Equation~\eqref{eq:jeff} follows from Jeffery's equation and encodes the reorientation dynamics of rod-like particles driven by the velocity gradient. The active stress $\bSigma^\ma$ in Eq.~\eqref{eq:actfl} is given by the volume-averaged particle stress arising from dipolar activity and is written as
\begin{equation}
    \bSigma^\ma = \alpha c(\bx) \bQ,
\end{equation}
where $c(\bx,t) = \int_{\mathbb{S}^1} \Psi \, \md\bp$ is the local concentration, $\bQ(\bx,t) = \frac{1}{c(\bx)}\int_{\mathbb{S}^1} \bp \bp \Psi \ \md \bp$ is the nematic tensor of the suspension, and $\alpha$ is the strength of the activity; in this paper, we restrict ourselves to $\alpha < 0$, corresponding to extensile stresses, as realized experimentally in reconstituted suspensions of MT-kinesin motor assemblies or in suspensions of pusher-type bacteria.
 
Evolving the high-dimensional distribution function $\Psi(\bx,\bp,t)$ is computationally expensive. To make the problem tractable, we instead reformulate the dynamics in terms of concentration $c(\bx,t)$ and the nematic tensor $\bQ(\bx,t)$, both of which are evolved using the interior suspension velocity field $\bu^\mi$; for notational convenience, we drop the superscript $`\mi'$ in the expressions below. The evolution equation for the concentration field is obtained by integrating Eq.~\eqref{eq:smoluchowski} over $\bp$
\begin{align}
  \partial_t c + \bu \cdot \nabla c &= d_{\mathrm{T}} \Delta c. \label{eq:concentration}
\end{align}
\begin{figure}[H]
    \centering  
    \includegraphics[width=0.4\linewidth]{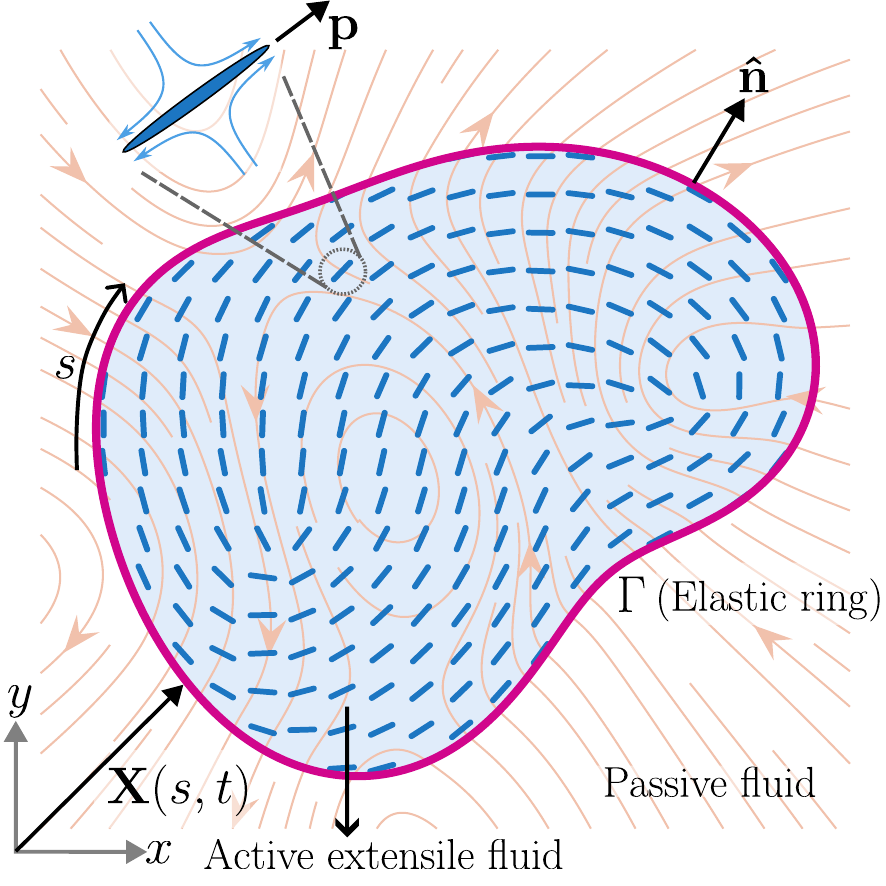}
    \caption{\textbf{Illustration of the conceived setup}. An elastic vesicle with bending and stretching elasticity confines a dilute suspension of extensile rods and is suspended in a Stokesian fluid of identical viscosity. The membrane $\Gamma$ is parameterized by the arc-length $s$, and any material point on it is identified as $\bX(s,t)$. The blue lines indicate the director field associated with the nematic tensor, and we show the associated streamlines in orange.}
    \label{fig:schematic}
\end{figure}
For an incompressible interior velocity field, any initial concentration fluctuations decay, and $c(\bx,t)$ approaches a spatially uniform value $c_0$ regardless of initial conditions. With $c$ pinned to this constant, $\bQ(\bx,t)$ remains the only relevant dynamical field in the problem. By taking orientational moments of Eq.~\eqref{eq:smoluchowski}, we obtain the evolution equation for $\bQ$ as 
\begin{equation}
   \overset{\nabla}{\mathbf{Q}} + 2 \mathbf{E} : \mathbf{S} = d_{\mathrm{T}} \Delta \mathbf{Q} - 4d_{\mathrm{R}} \left( \mathbf{Q} - \frac{\mathbf{I}}{2} \right). \label{eq:nematic_compact} 
\end{equation}
Here, we have introduced the upper-convected derivative $\overset{\nabla}{\bQ} = \partial_t \bQ + \bu \cdot \nabla \bQ - \left( \nabla \bu \cdot \bQ + \bQ \cdot \nabla \bu^T \right)$
and the fourth-order orientational moment $\mathbf{S} = \dfrac{1}{c_0} \int_{\mathbb{S}^1} \bp\bp\bp\bp\, \Psi \, \md\bp$. To close its evolution equation, we adopt the quadratic closure $\mathbf{S} \approx \bQ\bQ$ \cite{rautu2026}, yielding a closed system for the nematic tensor alone.

\subsection{Boundary conditions}

To solve for the interior and exterior velocity fields, we impose kinematic and dynamic boundary conditions on the vesicle interface $\Gamma$. The vesicle boundary is parametrized by its arc length $s$, and any material point is identified by a Lagrangian marker $\mathbf{X}(s,t)$, such that $\bx = \mathbf{X}(s,t)$ for $\bx \in \Gamma$ (see Fig.~\ref{fig:schematic}). The kinematic condition enforces continuity of velocity across the interface:
\begin{equation}\label{eq:kinematic}
    \bu^\me(\bx) = \bu^\mi(\bx), \quad \bx \in \Gamma.
\end{equation}
Since the membrane is a material surface, the Lagrangian points $\bX(s,t)$ are advected by this continuous velocity field,
\begin{equation}
\label{eq:membrane_advection}
    \frac{\partial \mathbf{X}(s, t)}{\partial t} = \mathbf{u}\left[\mathbf{X}(s,t)\right].
\end{equation}
The dynamic condition balances the jump in hydrodynamic stress across $\Gamma$ against the traction exerted by the vesicle
\begin{equation}\label{eq:dynamic}
    \left[\bSigma^\me(\bx) - \bSigma^\mT(\bx)\right] \cdot \uvc{n}(\bx) \equiv \llbracket \bSigma \rrbracket \cdot \uvc{n} = -\mathbf{F}[\mathbf{X}(s,t)], \quad \bx \in \Gamma,
\end{equation}
where $\uvc{n}$ is the unit normal on $\Gamma$, pointing into the passive exterior fluid, and $\bF[\bX(s,t)]$ is the elastic force density from the vesicle on the fluid. To simplify notation in what follows, we have also introduced a jump operator $\llbracket (.) \rrbracket \equiv (.)^\me - (.)^\mi$. Finally, the evolution of the nematic tensor $\bQ(\bx,t)$ needs specification of a boundary condition on $\Gamma$. In this work, we consider an anchoring boundary condition of the form
\begin{equation}\label{eq:anchoring}
    \uvc{n} \cdot \nabla \bQ = W \uvc{t}\uvc{t}, \quad \bx \in \Gamma,
\end{equation}
where $\uvc{t}$ is the unit tangent vector on the vesicle boundary, and $W$ is an anchoring strength. In the limit of $W \to 0$, one recovers the natural boundary condition for the nematic tensor, and as $W \to \infty$, we obtain parallel or tangential anchoring on the interface. In this paper, we will focus on a limit of $W \gg 1$, resulting in strong anchoring on the vesicle boundary. We now proceed to describe the mechanics of the vesicle ring, from which the elastic traction jump $\mathbf{F}$ is obtained, and detail how the above anchoring condition is approximately enforced via a weak penalty formulation.

\subsection{Vesicle mechanics and alignment interactions}
The vesicle boundary $\Gamma$ is parametrized by arc-length $s \in [0, L_0]$, where $L_0 = 2\pi R_0$ is the reference contour length of the undeformed circular vesicle. The unit tangent vector is then obtained as $\uvc{t}(s,t) = \bX_s / |\bX_s|$, which reduces to $\bX_s$ in the inextensible limit $|\bX_s| = 1$. The mechanical response of the vesicle is governed by the elastic free-energy functional
\begin{equation}
\label{eq:elastic_energy_functional}
    \mathcal{E}_{\mathrm{el}}[\bX] = \frac{1}{2} \int_{0}^{L_0}
    \left[ K_S \left( |\bX_s| - 1 \right)^2 + K_B |\bX_{ss}|^2
    \right] \, \mathrm{d}s,
\end{equation}
where $K_S$ is the stretching modulus, $K_B$ is the bending rigidity, and $|\bX_{ss}|$ is the local curvature of the boundary. The corresponding elastic force density of the vesicle, obtained by a functional derivative of $\mathcal{E}_\mathrm{el}$, is
\begin{equation}\label{eq:membrane_force}
    \mathbf{F}_{\mathrm{el}} = -\frac{\delta \mathcal{E}_{\mathrm{el}}}
    {\delta \bX} = -K_B \bX_{ssss} + K_S \partial_s
    \left[(|\bX_s| - 1)\, \uvc{t}\,\right],
\end{equation}
where the first term captures bending elasticity and the second captures stretching elasticity, with $K_S(|\bX_s|-1)$ playing the role of an internal line tension along the vesicle boundary. 

We now outline how the anchoring boundary condition on $\Gamma$ is enforced weakly through a penalty formulation. Following \cite{weady2024}, we model the anchoring as a steric interaction between the rod-like active particles and the vesicle by a Maier--Saupe-like potential, derived from a single interaction energy. Since the vesicle is a one-dimensional object, this interaction is carried on the curve $\Gamma$ itself and is written as a line integral along the boundary. The associated total interaction energy for the vesicle is given by
\begin{equation}\label{eq:fibal}
    \mathcal{E}_\mathrm{f}
    = -\frac{\beta_f}{2}\int_0^{L_0} \bQ\big|_{\bX} : \uvc{t}\,\uvc{t}\ \md s,
\end{equation}
where $\bQ|_{\bX}$ is the nematic tensor evaluated at the Lagrangian point $\bX(s,t)$ and $\beta_f > 0$ is the interaction strength with units of energy per unit length. Over trace-preserving nematic configurations, this energy is minimized when $\bQ = \uvc{t}\uvc{t}$ on $\Gamma$. Depending on the strength of $\beta_f$, the above energy provides an approximate enforcement of the boundary condition~\eqref{eq:anchoring}; the exact anchoring condition is recovered in the limit $\beta_f \to \infty$. Varying $\mathcal{E}_\mathrm{f}$ with respect to the ring configuration for a given suspension configuration yields the alignment force density on the ring as \cite{weady2024}
\begin{equation}\label{eq:fibalforce}
    \bF_\mathrm{f} = -\frac{\delta \mathcal{E}_\mathrm{f}}{\delta \bX}
    = -\beta_f\, \partial_s \left[(\bI - \uvc{t}\uvc{t}) \cdot
      \bQ\big|_{\bX} \cdot \uvc{t}\right].
\end{equation}
The above expression, together with the passive elastic force density in Eq.~\eqref{eq:membrane_force}, provides the total traction jump $\bF = \bF_\mathrm{el} + \bF_\mathrm{f}$ across $\Gamma$ in Eq.~\eqref{eq:dynamic}.

By Newton's third law, the same steric-alignment interaction reacts on the suspension as a torque on the particles. To express this reaction in the Eulerian frame, we introduce the alignment tensor field carried by the ring,
\begin{equation}\label{eq:Tfield}
    \boldsymbol{\mathcal{T}}(\bx,t)
    = \int_0^{L_0} \uvc{t}\,\uvc{t}\;
      \delta\big(\bx - \bX(s,t)\big)\ \md s
    \;\equiv\; \uvc{t}\,\uvc{t}\;\delta_\Gamma(\bx),
\end{equation}
where $\delta_\Gamma(\bx)$ is the line delta function supported on $\Gamma$. The field $\boldsymbol{\mathcal{T}}$ is the natural Eulerian representation of the membrane tangent: it is a surface density rather than a bulk field, reflecting the fact that the alignment interaction is strictly local to the ring. A particle at position $\bx$ with orientation $\bp$ then experiences the orientational potential $\mathcal{U}_\mathrm{p} = -\tfrac{1}{2}\tilde{\beta}\, \bp\bp : \boldsymbol{\mathcal{T}}(\bx)$, whose torque $-\nabla_\bp\, \mathcal{U}_\mathrm{p}$ modifies the rotational flux in Eq.~\eqref{eq:jeff} to
\begin{equation}\label{eq:p_dynamics_alignment}
    \dot{\bp} = (\bI - \bp\bp) \cdot
    \big(\nabla \bu + \beta_p\, \boldsymbol{\mathcal{T}}\big) \cdot \bp
    - d_\mathrm{R}\, \nabla_\bp \ln \Psi,
\end{equation}
where $\beta_p$ is the microscopic energy scale $\tilde{\beta}$ divided by the rotational drag of a rod and carries dimensions of length over time. Integrating $\mathcal{U}_\mathrm{p}$ against the particle distribution $\Psi(\bx,\bp,t)$ recovers Eq.~\eqref{eq:fibal} with $\beta_f = c_0\tilde{\beta}$: the alignment traction on the vesicle \eqref{eq:fibalforce} and the particle torque in \eqref{eq:p_dynamics_alignment} are the action--reaction pair of the same coupling, so the model contains a single independent alignment strength rather than two adjustable parameters. Taking orientational moments of the kinetic equation with the modified flux, the evolution of the nematic tensor becomes
\begin{equation}\label{eq:Qev}
    \overset{\nabla}{\bQ} + 2\,\bQ(\bQ : \mathbf{E})
    = \beta_p \left[\boldsymbol{\mathcal{T}} \cdot \bQ
      + \bQ \cdot \boldsymbol{\mathcal{T}}
      - 2\,\bQ(\bQ : \boldsymbol{\mathcal{T}})\right]
    + d_\mathrm{T} \Delta \bQ
    - 4 d_\mathrm{R} \left( \bQ - \frac{\bI}{2} \right),
\end{equation}
where we note that the alignment term is linear in $\boldsymbol{\mathcal{T}}$ and away from $\Gamma$ the bulk orientational dynamics are strictly unaltered. 

In summary, the two faces of this single energetic coupling constitute a bidirectional interaction: the torque in Eq.~\eqref{eq:p_dynamics_alignment} rotates the particles toward the local membrane tangent, while the traction in Eq.~\eqref{eq:fibalforce} pulls the membrane tangent toward the local nematic axis. Taking $\{\beta_p, \beta_f\} < 0$ yields normal (homeotropic) anchoring instead. The alignment potential also induces a constraint stress in the suspension, which we neglect in the dilute limit considered here \cite{weady2024}. As we show in Sec.~\ref{sec:lsa}, the surface-delta structure of $\boldsymbol{\mathcal{T}}$ permits an exact reduction of these alignment interactions to an effective boundary condition on $\bQ$, which also makes the connection to the anchoring strength $W$ in Eq.~\eqref{eq:anchoring} precise.

\subsection{Nondimensionalization}
The equations of motion are now non-dimensionalized using the unperturbed ring radius $l^* = R_0$ as the length scale and the translational diffusion time $t^* = R_0^2/d_T$ as the time scale. The distribution function $\Psi$ is normalized such that $\int_{\Omega^\mi}\int_{\mathbb{S}^1} \Psi\ \md\bp\, \md\bx = |\Omega^\mi|$, which sets $c_0 = \mathcal{N}/|\Omega^\mi| = 1$. The resulting dimensionless groups are summarized in Table~\ref{tab:dimensionless_params}. We note that the two alignment strengths scale differently with the problem variables: $\beta_p$ carries units of length per time, as it multiplies the surface density $\boldsymbol{\mathcal{T}}$ in Eq.~\eqref{eq:Qev}, while $\beta_f$ has units of energy per unit length. Since both descend from the same microscopic steric interaction, their dimensionless counterparts are not independent; following \cite{weady2024}, we set them equal and denote the single dimensionless alignment strength by $\beta \equiv t^*\beta_p/l^* = t^*\beta_f/\mu$.
\begin{table}[H]
\begin{ruledtabular}
\begin{tabular}{lcc}
\textbf{Physical Parameter} & \textbf{Symbol} & \textbf{Scaling Expression} \\
\hline
Activity strength & $\alpha$ & $(t^*/\mu)\,\alpha$ \\
Rotational diffusivity & $d_R$ & $t^*\, d_R$ \\
Stretching modulus & $K_S$ & $(t^*/\mu)\,K_S$ \\
Bending rigidity & $K_B$ & $(t^*/(\mu\, l^{*2}))\,K_B$ \\
Alignment strength & $\beta$ & $(t^*/l^*)\,\beta_p = (t^*/\mu)\,\beta_f$ \\
\end{tabular}
\caption{Dimensionless parameters of the active vesicle model. \label{tab:dimensionless_params}}
\end{ruledtabular}
\end{table}
In these units, and retaining the same symbols for the dimensionless variables, the evolution of the nematic tensor becomes
\begin{equation}
\label{eq:dimensionless_nematic_evolution}
    \overset{\nabla}{\bQ} + 2\,\bQ(\bQ : \mathbf{E})
    = \beta \left[\boldsymbol{\mathcal{T}} \cdot \bQ
      + \bQ \cdot \boldsymbol{\mathcal{T}}
      - 2\,\bQ(\bQ : \boldsymbol{\mathcal{T}})\right]
    + \Delta \bQ
    - 4 d_R \left( \bQ - \frac{\bI}{2} \right),
\end{equation}
and the traction jump across the ring is given by
\begin{equation}
\label{eq:dimensionless_force}
    \bF = -K_B\, \bX_{ssss}
    + K_S\, \partial_s \left[ (|\bX_s| - 1)\, \uvc{t} \right]
    - \beta\, \partial_s \left[ (\bI - \uvc{t}\uvc{t}) \cdot
      \bQ\big|_{\bX} \cdot \uvc{t} \right].
\end{equation}
To isolate the core mechanisms driving the vesicle dynamics, we constrain several of these parameters. The stretching modulus $K_S$ is set sufficiently high that the membrane is effectively inextensible, and the alignment strength $\beta$ is chosen large enough to enforce robust tangential order at the boundary. With these auxiliary parameters held fixed, the system is governed by two primary free parameters, the activity strength $\alpha$ and the bending rigidity $K_B$, and our numerical and analytical investigations map the vesicle dynamics across this two-dimensional $(\alpha, K_B)$ parameter space.

\subsection{Computational method}\label{sec:numerics}
Simulating the full FSI problem requires solving a coupled system of nonlinear partial differential equations on a time-evolving domain: a significant computational challenge. We address this using the Immersed Boundary Method (IBM)~\cite{peskin2002}, a hybrid Eulerian--Lagrangian framework in which the fluid equations are solved on a fixed Cartesian grid while the vesicle boundary is tracked as a set of moving Lagrangian material points. All simulations are performed in a doubly-periodic domain of size $L \times L$, discretized by $N$ grid points per direction with uniform spacing $h = L/N$. The vesicle boundary is represented by $N_b$ Lagrangian markers $\bX_k(t)$, $k = 1,\dots,N_b$, with arc-length spacing $\Delta s = 2\pi/N_b$ in the undeformed state. In all nonlinear simulations, we take $N \approx 256\text{--}512$ and choose $N_b$ to maintain $\Delta s \approx h/2$, which minimizes numerical fluid leakage through the vesicle boundary.

The IBM couples the Lagrangian and Eulerian frames through spreading and interpolation operators built from a regularized delta function $\delta_h$, for which we use the Peskin 4-point kernel throughout~\cite{peskin2002}. The traction jump $\bF[\bX(s,t)]$ derived in the previous section is spread from the Lagrangian markers onto the Eulerian grid as a body force,
\begin{equation}
\label{eq:force_spreading}
    \bff(\bx, t) = \sum_{k=1}^{N_b} \bF_k(t)\,
    \delta_h(\bx - \bX_k(t))\,\Delta s,
\end{equation}
which approximately enforces the stress jump condition~\eqref{eq:dynamic} across $\Gamma$ without an explicit treatment of the interface. The same construction regularizes the alignment field $\boldsymbol{\mathcal{T}}$ of Eq.~\eqref{eq:Tfield}: the line delta $\delta_\Gamma$ is replaced by $\delta_h$, and the tangent tensor is spread as
\begin{equation}
\label{eq:T_spreading}
    \boldsymbol{\mathcal{T}}(\bx, t) = \sum_{k=1}^{N_b}
    \uvc{t}_k(t)\,\uvc{t}_k(t)\,
    \delta_h(\bx - \bX_k(t))\,\Delta s.
\end{equation}
Because the alignment terms in Eq.~\eqref{eq:dimensionless_nematic_evolution} are linear in
$\boldsymbol{\mathcal{T}}$, and the kernel $\delta_h$ has unit integral by construction, the total alignment torque exerted on the suspension is independent of the grid resolution; the regularization sets the width of the anchoring layer but not its integrated strength. Finally, the kinematic condition is enforced by interpolating the Eulerian velocity back onto the Lagrangian markers,
\begin{equation}
\label{eq:velocity_interpolation}
    \frac{\partial \bX_k}{\partial t} = \sum_{\bx}
    \bu(\bx, t)\,\delta_h(\bx - \bX_k(t))\,h^2,
\end{equation}
where the sum runs over all grid points. Using the same kernel for spreading and interpolation renders the two operators discretely adjoint, ensuring that the coupling conserves energy in the semi-discrete sense.

A key advantage of the IBM is that the interior and exterior hydrodynamic problems are solved simultaneously on the same fixed grid. To this end, we introduce an indicator function $\chi(\bx,t)$ that strictly confines the active stress to the vesicle interior and evaluates to unity inside $\Gamma$ and zero outside. To prevent numerical instabilities and non-physical sharp discontinuities at the fluid-membrane interface, the indicator function is regularized by a convolution with a Gaussian filter kernel $G_\sigma$ as $\chi_s(\bx,t) = \chi(\bx,t) * G_\sigma$. This smoothed indicator function ensures the active stress decays smoothly to zero outside the vesicle and recasts the two-domain problem of Sec.~\ref{sec:model} as a single momentum balance over the whole domain as
\begin{equation}
\label{eq:stokes_unified}
    -\nabla q + \Delta \bu +
    \nabla \cdot (\chi_s\, \bSigma^\ma) + \bff = \mathbf{0},
    \quad \nabla \cdot \bu = 0.
\end{equation}
Given the periodic domain, Eq.~\eqref{eq:stokes_unified} is solved spectrally at each time step. A known difficulty of this approach is that the discrete gradient and divergence operators are not adjoint to the spreading and interpolation kernels, leading to slow fluid leakage out of the enclosed vesicle. We correct for this using the modified spectral operators as proposed in~\cite{peskinprintz1993}, which enforce exact discrete incompressibility and eliminate the spurious volume loss.

The overall time-stepping proceeds as follows. The nematic tensor $\bQ$ is evolved with an implicit-explicit (IMEX) scheme, treating the diffusive term implicitly and the nonlinear advective and alignment terms explicitly. The Stokes problem~\eqref{eq:stokes_unified} is quasi-static and is solved at each step given the current $\bQ$ and $\bX_k$. The membrane markers are then advanced by a forward Euler step, $\bX_k^{n+1} = \bX_k^n + \Delta t\, \bU_k^n$, where $\bU_k^n$ is the divergence-free velocity interpolated via Eq.~\eqref{eq:velocity_interpolation}.

\section{Linear Stability Analysis}\label{sec:lsa}

We will illustrate in Sec.~\ref{sec:nonlinear} that the nonlinear simulations described above reveal a sequence of morphological transitions as activity is increased. To understand the mechanism of such morphodynamics, we first perform a linear stability analysis (LSA) of the equilibrium configuration of the vesicle; for the sake of simplicity, we set the rotational diffusion $d_R = 0$. The base state of our problem is a circular vesicle of unit radius, $R = 1$ in dimensionless units, with a quiescent flow $\bu = \mathbf{0}$. Activity generates no motion in this configuration, and its signature is instead carried by the pressure field and the structure of the confined nematic. We first outline how we can obtain this base state configuration. 

\subsection{Base state and boundary conditions for the nematic tensor}

In the absence of rotational diffusion, the steady-state equation for the nematic tensor from Eq.~\eqref{eq:dimensionless_nematic_evolution} reduces to
\begin{equation}\label{eq:baseQeq}
    \Delta \bQ + \beta\, \delta_\Gamma\, \mathcal{A}[\bQ] = \mathbf{0},
    \ \ \text{with} \ \ 
    \mathcal{A}[\bQ] \equiv \uvc{t}\uvc{t} \cdot \bQ
    + \bQ \cdot \uvc{t}\uvc{t} - 2\,\bQ(\bQ : \uvc{t}\uvc{t}),
\end{equation}
where we have used the definition of the alignment field $\boldsymbol{\mathcal{T}} = \uvc{t}\uvc{t}\,\delta_\Gamma$ and the linearity of the alignment terms in $\boldsymbol{\mathcal{T}}$ to factor out the line delta function. Away from $\Gamma$, the nematic tensor is thus harmonic, and the entire alignment interaction acts as a singular line source concentrated on the membrane. This structure permits an exact reduction of the anchoring interaction to a boundary condition. Integrating Eq.~\eqref{eq:baseQeq} over a pillbox of vanishing thickness straddling the interface, the harmonic contribution reduces to the difference of normal fluxes on the two faces, while the delta function integrates to unity, yielding a jump condition
\begin{equation}\label{eq:fluxjump}
    \uvc{n} \cdot \nabla \bQ \Big|_{\Gamma^+}
    - \uvc{n} \cdot \nabla \bQ \Big|_{\Gamma^-}
    = -\beta\, \mathcal{A}[\bQ]\Big|_{\Gamma},
\end{equation}
where $\Gamma^\pm$ denote the exterior and interior faces of the interface. In our numerical scheme, the nematic tensor is evolved over the entire periodic domain, so the alignment torque is shared between the fields on either side of the interface. For harmonic fields inside and outside a circle, the two sides carry equal flux, and the jump splits evenly between them; the interior field, which alone carries physical meaning, therefore satisfies the Robin boundary condition
\begin{equation}\label{eq:robin}
    \uvc{n} \cdot \nabla \bQ\Big|_{\Gamma} = W \mathcal{A}[\bQ]\Big|_\Gamma, \ \ \text{with} \ \ W = \frac{\beta}{2}.
\end{equation}
The above expression makes the connection to the boundary condition in Eq.~\eqref{eq:anchoring} explicit; as $W \to \infty$, the condition $\mathcal{A}[\bQ] = \mathbf{0}$ enforces sharp tangential anchoring, $\bQ = \uvc{t}\uvc{t}$ on $\Gamma$. The length scale $1/W$ plays the role of the extrapolation length familiar from weak anchoring of liquid crystals at surfaces~\cite{chandler2024}. The problem for the base state is then simply
\begin{equation}
    \Delta \bQ = \mathbf{0} \quad \text{in } \Omega^\mi,
\end{equation}
supplemented with the Robin condition~\eqref{eq:robin} on $\Gamma$.

\begin{figure}[t!]
    \centering
    \includegraphics[width=1\textwidth]{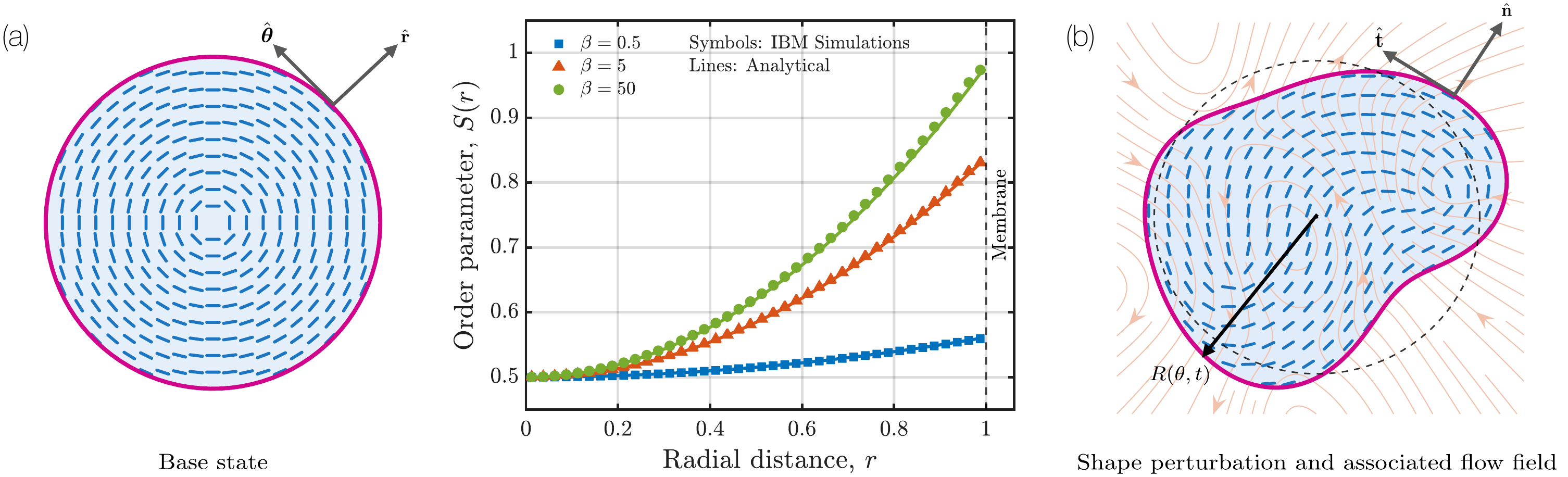}
    \caption{\textbf{Base state and perturbation.} (a) The director field associated with the $+1$ vortex of the active suspension in the circular base state of the vesicle. The nematic order parameter $S(r)$ grows quadratically with distance from the center and reaches its maximum at the vesicle interface. The numerical solution (symbols) from the IBM agrees well with the analytical solution (lines) obtained using Robin boundary conditions. Increasing the alignment strength $\beta$ increases the tangential order and thus $S(1)$  at the interface. (b) Perturbation to the circle with the associated coordinate system.}
    \label{fig:LSA}
\end{figure}

For the unperturbed circular vesicle, the boundary condition is axisymmetric, and we propose a trace-preserving Ansatz
\begin{equation}
    \bQ = \frac{\bI}{2} + \varpi(r) \left[\uvc{r}\uvc{r} -
    \uvc{\theta}\uvc{\theta}\right],
\end{equation}
where $\varpi(r)$ measures the local degree of nematic order: $\varpi = 0$ corresponds to an isotropic suspension, and $\varpi = \mp 1/2$ to perfect tangential (radial) alignment. Harmonicity of $\bQ$ requires
\begin{equation}
    \varpi''(r) + \frac{1}{r}\varpi'(r) - \frac{4}{r^2} \varpi(r) = 0,
\end{equation}
whose solutions are $\varpi(r) = a r^2 + b r^{-2}$; regularity at the origin sets $b = 0$. The amplitude $a$ follows from the Robin condition, which yields
\begin{equation}
    \partial_r \bQ \Big|_{r=1} = 2 W \left(\frac{1}{4} - a^2\right)
    \left(\uvc{\theta} \uvc{\theta} - \uvc{r} \uvc{r}\right),
\end{equation}
and matching coefficients yields
\begin{equation}\label{eq:aofW}
    a = -\frac{\sqrt{1+W^2} - 1}{2 W}.
\end{equation}
Taken together, the base-state nematic tensor is given by
\begin{equation}
    \bQ^{(0)} = \frac{1 + 2 a r^2}{2}\, \uvc{r} \uvc{r}
    + \frac{1 - 2 a r^2}{2}\, \uvc{\theta}\uvc{\theta}.
\end{equation}
The physical picture is as follows. At the center, the nematic tensor is
isotropic, $\bQ^{(0)}(0) = \bI/2$. The base state is thus a $+1$ vortex with a topological defect at the origin, as required by the tangential anchoring condition on a disk. Moving outward, the anchoring propagates into the bulk by translational diffusion, and the degree of tangential order grows quadratically with radius, reaching its maximum $|a|$ at the interface. The anchoring strength controls this amplitude through Eq.~\eqref{eq:aofW}: $a \simeq -W/4$ for weak anchoring, while $a \to -1/2$ as $W \to \infty$, where the boundary attains perfect tangential order. Notably, $a(W)$ saturates rapidly, and at $W = 4$ the boundary order reaches $80\%$ of its sharp-anchoring value; the base state thus becomes insensitive to the precise anchoring strength once $W \gtrsim 1$. We have verified this solution against full numerical simulations of the base state as shown in Fig.~\ref{fig:LSA}; the measured nematic order $S=\left(2s-1\right)/2$,
where $s$ is the largest eigenvalue of $\bQ$, follows the predicted $r^2$ profile with amplitude given by Eq.~\eqref{eq:aofW} at $W = \beta/2$, with no fitting parameters. 

\subsection{Stress balance in the base state}

In the quiescent base state, the interior stress carries only pressure and the active contribution from the nematic order, with the total stress given by
\begin{equation}
    \bSigma^{\mT,(0)} = -q^{\mi,(0)}\, \bI + \alpha\, \bQ^{(0)}.
\end{equation}
With $\bu = \mathbf{0}$, momentum balance demands
\begin{equation}
    \nabla q^{\mi,(0)} = \alpha\, \nabla \cdot \bQ^{(0)}
    \implies q^{\mi,(0)}(r) = 2 a \alpha\, r^2,
\end{equation}
where the pressure gauge is set to zero at the defect core without any loss of generality; the radially growing nematic order of the $+1$ vortex is thus
accompanied by a parabolic pressure profile. Outside, the fluid is stress-free up to a constant pressure, $q^{\me,(0)} = q_0$, which is fixed by
the dynamic boundary condition~\eqref{eq:dynamic}. For the circular vesicle, only bending contributes to the elastic force density, $\bF_0 = -K_B\,\uvc{r}$, and the normal stress balance reads
\begin{equation}
   \llbracket \bSigma^{(0)} \rrbracket \cdot \uvc{r}
   = -q_0\, \uvc{r} - \bSigma^{\mT,(0)}(1) \cdot \uvc{r}
   = -\bF_0 \implies q_0 = -K_B + \frac{\alpha}{2}\,(2a - 1).
\end{equation}
The base state is therefore an equilibrium in which active stresses are
balanced jointly by the pressure fields and the bending stresses of the ring. For the domain perturbation that follows, we will require the stress jump on the interface and its radial gradient; using the above results, we obtain
\begin{align}
    \llbracket \bSigma^{(0)} \rrbracket \Big|_{r=1}
    &= K_B\, \bI + 2 a \alpha\, \uvc{\theta}\uvc{\theta}, \\
    \partial_r \llbracket \bSigma^{(0)} \rrbracket \Big|_{r=1}
    &= 2 a \alpha \left[\uvc{r}\uvc{r} + 3\, \uvc{\theta}\uvc{\theta}\right],
\end{align}
where the second relation follows from the constancy of the exterior stress.
\subsection{Domain perturbation Ansatz}

To probe the stability of this equilibrium, we now perturb the membrane as $R(\theta,t) = 1 + \epsilon\, \xi(\theta,t)$, where $\epsilon \ll 1$ and $\xi(\theta,t)$ is decomposed into azimuthal Fourier modes,
\begin{equation}\label{eq:radial_perturbation}
    \xi(\theta, t) = \sum_m \zeta_m\, e^{\mi m \theta + \sigma_m t},
    \qquad m \in \mathbb{Z},
\end{equation}
with modal amplitudes $\zeta_m$ and unknown growth rate $\sigma_m = \sigma(m)$. At linear order, the modes evolve independently, so it suffices to study one mode at a time. We therefore fix $m$ and suppress the subscript on all modal amplitudes for the sake of notational simplicity. We now write every perturbation field as a radial profile multiplying the common phase factor $\Phi_m \equiv e^{\mi m \theta + \sigma t}$; thus $\xi(\theta,t) = \zeta\,\Phi_m$, with $\zeta$ now a constant amplitude, $\sigma$ being the growth rate of the mode, and the summation over modes is implied.

The unit tangent and outward normal are perturbed away from their base-state values $\uvc{\theta}$ and $\uvc{r}$ in this deformed configuration. To leading order in $\epsilon$,
\begin{align}\label{eq:linearized_tangent}
    \uvc{t} &\approx \uvc{\theta} + \epsilon\, \mi m\, \zeta\, \Phi_m\, \uvc{r}, \\
\label{eq:linearized_normal}
    \uvc{n} &\approx \uvc{r} - \epsilon\, \mi m\, \zeta\, \Phi_m\, \uvc{\theta},
\end{align}
where the normal follows from $\uvc{t}\cdot\uvc{n} = 0$ to $\OR(\epsilon^2)$. The nematic tensor is perturbed about its base state as $\bQ = \bQ^{(0)} + \epsilon\, \tilde{\bQ}$. Symmetry and preservation of the unit trace restrict the perturbation to two independent components, which in the polar basis is written as
\begin{equation}
\label{eq:perturbed_nematic_order}
    \tilde{\bQ}(r, \theta, t) =
    \begin{bmatrix} A(r) & B(r) \\ B(r) & -A(r) \end{bmatrix} \Phi_m,
\end{equation}
where $A(r)$ and $B(r)$ are the radial profiles of the normal and shear components of the orientational perturbation. Because the domain itself deforms, the value of $\bQ$ on the interface differs from its perturbation at $r = 1$. As is classical in domain perturbation problems, we Taylor expand the variables about the undeformed circle to obtain
\begin{equation}\label{eq:Q_at_interface}
    \bQ(R,\theta) = \bQ^{(0)}\big|_{r=1}
    + \epsilon \left[\xi\, \partial_r \bQ^{(0)}\big|_{r=1}
    + \tilde{\bQ}(1,\theta,t)\right]
    = \begin{bmatrix} \tfrac{1}{2}+a & 0 \\[2pt] 0 & \tfrac{1}{2}-a \end{bmatrix}
    + \epsilon \begin{bmatrix}
        2a\zeta + A(1) & B(1) \\[2pt] B(1) & -2a\zeta - A(1)
    \end{bmatrix} \Phi_m,
\end{equation}
which shows that a shape deformation alone, even at fixed $\tilde{\bQ}$,
alters the nematic order sampled by the vesicle interface. The stress fields are perturbed identically as
$\bSigma^{\me,\mT} = \bSigma^{\me,\mT,(0)}(r)
+ \epsilon\, \tilde{\bSigma}^{\me,\mT}$, with
\begin{equation}
\label{eq:perturbed_stress}
    \tilde{\bSigma}^{\me,\mT}(r,\theta,t) =
    \begin{bmatrix}
        \tilde\Sigma^{\me,\mT}_{rr}(r) & \tilde\Sigma^{\me,\mT}_{r\theta}(r) \\[2pt]
        \tilde\Sigma^{\me,\mT}_{r\theta}(r) & \tilde\Sigma^{\me,\mT}_{\theta\theta}(r)
    \end{bmatrix} \Phi_m,
\end{equation}
whose radial profiles are determined below by the solution of the associated Stokes problems. The triplet $\{\zeta, A(r), B(r)\}$ thus carries the essential dynamical content of the problem, and its components are coupled in a closed loop: a deformation of the interface forces the nematic field through the anchoring condition~\eqref{eq:robin}, evaluated on the perturbed geometry; in turn, the orientational perturbation $\{A, B\}$ generates active stresses in the bulk while the deformed ring exerts elastic and alignment tractions on the interface, and together these drive the perturbation flow; the flow then closes the loop by advecting the interface through the kinematic condition~\eqref{eq:kinematic}. Determining it requires making the linearized boundary conditions explicit and formulating the hydrodynamic problem, to which we turn next.
\subsection{Stress balance at the interface}

We first construct the interfacial force density $\bF$ of the perturbed
vesicle. To $\OR(\epsilon)$ the arc-length element of the deformed boundary is $\md s = (1 + \epsilon\,\zeta\,\Phi_m)\,\md\theta$, and for $m \geq 2$ the perturbation preserves the total perimeter at this order since
$\oint \xi\, \md\theta = 0$. For perimeter-preserving modes ($m \geq 2$), the deformation is compatible with an $\OR(\epsilon)$ tangential redistribution of material points that maintains local inextensibility, $|\bX_s| - 1 \sim \OR(\epsilon^2)$. We therefore neglect the stretching contribution to the force density; the bending contribution, evaluated on the perturbed shape, is
\begin{equation}\label{eq:ellin}
   \bF_{\mathrm{el}} = -K_B\, \partial_s^4 \bX
   = -K_B\, \uvc{r}
   + \epsilon\, K_B \left[-(m^2-1)(m^2+3)\,\zeta\, \uvc{r}
   + \mi m (3m^2-2)\,\zeta\, \uvc{\theta}\right] \Phi_m
   + \OR(\epsilon^2),
\end{equation}
where the base-state term is the uniform bending force of the circle. The
elastic traction associated with the alignment interactions is evaluated using the boundary value of the nematic tensor~\eqref{eq:Q_at_interface} together with the rotated tangent and normal, Eqs.~\eqref{eq:linearized_tangent}--\eqref{eq:linearized_normal}. On noting
$\bI - \uvc{t}\uvc{t} = \uvc{n}\uvc{n}$, we find
$\uvc{n}\cdot\bQ\cdot\uvc{t} = \epsilon\,[B(1) + 2\mi a m \zeta]\,\Phi_m$,
where the second term arises from the rotation of the boundary frame
against the ordered base state. Using these relations for the alignment force density as defined in Eq.~\eqref{eq:fibal}, we obtain
\begin{equation}\label{eq:flin}
    \bF_{\mathrm{f}}
    = -\beta\, \partial_s \left[(\bI - \uvc{t}\uvc{t}) \cdot \bQ\big|_{\bX}
    \cdot \uvc{t}\right]
    = \epsilon\, \beta \left[2 a m \zeta - \mi B(1)\right]
    \left(m\, \uvc{r} - \mi\, \uvc{\theta}\right) \Phi_m
    + \OR(\epsilon^2).
\end{equation}
We note that even an undistorted nematic field ($B = 0$) thus exerts an alignment traction on a deformed boundary, and conversely a shear distortion of the director field stresses the undeformed ring, reflecting the bidirectional coupling as discussed in Sec.~\ref{sec:model}. Collecting Eqs.~\eqref{eq:ellin}--\eqref{eq:flin}, the total force density can be written  as
\begin{equation}
    \bF = -K_B\, \uvc{r}
    + \epsilon \left[f_r\, \uvc{r} + f_\theta\, \uvc{\theta}\right]\Phi_m,
\end{equation}
with
\begin{align}
    f_r &= -K_B(m^2-1)(m^2+3)\,\zeta
          + \beta m \left[2am\zeta - \mi B(1)\right], \\
    f_\theta &= \mi m K_B(3m^2-2)\,\zeta
          - \mi \beta \left[2am\zeta - \mi B(1)\right].
\end{align}
This elastic force density balances the hydrodynamic traction jump across the vesicle. Taylor expanding the jump in the stress tensor $\llbracket\bSigma\rrbracket$ around the undeformed interface, we obtain
\begin{equation}
 \llbracket\bSigma\rrbracket\big|_{r=R}
 = \llbracket\bSigma^{(0)}\rrbracket\big|_{r=1}
 + \epsilon \left[\zeta\, \Phi_m\,
 \partial_r \llbracket\bSigma^{(0)}\rrbracket\big|_{r=1}
 + \llbracket\tilde{\bSigma}\rrbracket\big|_{r=1}\right]
 + \OR(\epsilon^2).
\end{equation}
Upon contracting with the perturbed normal~\eqref{eq:linearized_normal}, the traction jump becomes
\begin{equation}
  \llbracket\bSigma\rrbracket \cdot \uvc{n}
  = K_B\, \uvc{r} + \epsilon \left[
  - \mi m \zeta\, \llbracket\bSigma^{(0)}\rrbracket \cdot \uvc{\theta}\, \Phi_m
  + \zeta\, \partial_r \llbracket\bSigma^{(0)}\rrbracket \cdot \uvc{r}\, \Phi_m
  + \llbracket\tilde{\bSigma}\rrbracket \cdot \uvc{r}\right].
\end{equation}
The dynamic boundary condition~\eqref{eq:dynamic} demands
$\llbracket\bSigma\rrbracket \cdot \uvc{n} = -\bF$;  using the base-state jump and its gradient, the normal and
tangential projections provide us the following relations at $\mathcal{O}(\epsilon)$,
\begin{align}
\label{eq:normal_bc}
\llbracket\tilde\Sigma_{rr}\rrbracket\big|_{r=1} &= -f_r - 2 a \alpha\, \zeta, \\
\label{eq:shear_bc}
\llbracket\tilde\Sigma_{r\theta}\rrbracket\big|_{r=1} &=
    -f_\theta + \mi m \zeta\, (K_B + 2 a \alpha).
\end{align}
These two conditions encode how the deformed elastic ring forces the
surrounding fluid; together with the kinematic and anchoring conditions, they close the hydrodynamic problem, to which we turn next.

\subsection{Solution to the hydrodynamic problem}
\label{sec:stokes_solution}

The $\mathcal{O}(\epsilon)$ perturbed flow on either side of the interface is conveniently described by a stream function $\psi(r,\theta,t)$,
\begin{equation}
\label{eq:streamfunction_def}
    u_r = \frac{1}{r}\,\partial_\theta\psi, \qquad
    u_\theta = -\partial_r\psi,
\end{equation}
which satisfies incompressibility identically. Adopting the modal form
$\psi = \psi(r)\,\Phi_m$, the velocity components are
$\tilde u_r = (\mi m/r)\,\psi$ and $\tilde u_\theta = -\psi'$, where, in
keeping with our convention, we suppress the mode index on the radial
profile and write $\psi^{\mi}$ and $\psi^{\me}$ for the interior and
exterior solutions. Anticipating the boundary conditions, we first express
the perturbed stresses in terms of $\psi$. The shear component follows
directly from Eq.~\eqref{eq:streamfunction_def},
\begin{equation}
\label{eq:sigma_rtheta_psi}
    \tilde\Sigma_{r\theta}(r) = -\psi''
        + \frac{\psi'}{r} - \frac{m^2\psi}{r^2}
        + \chi\, \alpha B(r),
\end{equation}
while the normal component involves the pressure, which we eliminate using
the azimuthal momentum balance to obtain
\begin{equation}
\label{eq:sigma_rr_psi}
    \tilde\Sigma_{rr}(r) = \frac{\mi}{m}\left[
        -r\psi''' - \psi'' + \frac{(1+3m^2)\,\psi'}{r}
        - \frac{4m^2\psi}{r^2}
        + \chi\, \alpha\left(r B' + 2 B - 2 \mi m A\right)\right].
\end{equation}
Both expressions hold on either side of the interface, with the active
contribution, marked by the indicator function $\chi$, present only in the
interior problem. 

We solve the exterior problem first. The unforced stream function satisfies
the biharmonic equation $\Delta^2\psi^\me = 0$, which for mode $m$ reduces
to the equidimensional form
\begin{equation}
\label{eq:biharmonic_ode}
    \mathcal{S}_m[\psi] \equiv
    \frac{\md^4\psi}{\md r^4} + \frac{2}{r}\frac{\md^3\psi}{\md r^3}
    - \frac{2m^2+1}{r^2}\frac{\md^2\psi}{\md r^2}
    + \frac{2m^2+1}{r^3}\frac{\md\psi}{\md r}
    + \frac{m^4-4m^2}{r^4}\,\psi = 0,
\end{equation}
with the four solutions $r^{\pm m}$ and $r^{\pm m + 2}$. Imposing decay as
$r\to\infty$ retains
\begin{equation}
\label{eq:perturbed_velocity_potential}
    \psi^\me(r) = c\,r^{-m} + d\,r^{-m+2},
\end{equation}
with $\{c,d\}$ as unknown constants; the harmonic solution $r^{-m}$ carries no
pressure perturbations, while $r^{-m+2}$ carries the exterior pressure perturbation
$\tilde q^{\me} = 4\mi(m-1)\,d\,r^{-m}$. Evaluating
Eqs.~\eqref{eq:sigma_rtheta_psi}--\eqref{eq:sigma_rr_psi} on this solution,
the exterior tractions at the unperturbed interface are obtained as
\begin{align}
\label{eq:ext_traction_rr}
    \tilde\Sigma^\me_{rr}(1) &= -2\mi m(m+1)\,c - 2\mi(m-1)(m+2)\,d, \\
\label{eq:ext_traction_rtheta}
    \tilde\Sigma^\me_{r\theta}(1) &= -2m(m+1)\,c - 2m(m-1)\,d.
\end{align}
We can now eliminate the exterior flow in favor of the interior one. Continuity of velocity across the interface requires $\psi^\me(1) = \psi^\mi(1)$ and $\psi^{\me\prime}(1) = \psi^{\mi\prime}(1)$, which allows us to write
\begin{equation}
\label{eq:BD_from_continuity}
    \begin{bmatrix} c \\ d \end{bmatrix}
    = \frac{1}{2}
    \begin{bmatrix} 2-m & -1 \\ m & 1 \end{bmatrix}
    \begin{bmatrix} \psi^\mi(1) \\ \psi^{\mi\prime}(1) \end{bmatrix}.
\end{equation}
Substituted into
Eqs.~\eqref{eq:ext_traction_rr}--\eqref{eq:ext_traction_rtheta}, this
furnishes a Dirichlet-to-Neumann map: the exterior tractions are expressed
purely in terms of interior boundary data. The dynamic boundary conditions
then close the interior problem, taking the Robin-type form
\begin{equation}
\label{eq:sigma_star}
    \tilde\Sigma^\me_{rr}(1)
    - \tilde\Sigma^{\mT}_{rr}(1)
    = \llbracket\tilde\Sigma_{rr}\rrbracket, \qquad
    \tilde\Sigma^\me_{r\theta}(1)
    - \tilde\Sigma^{\mT}_{r\theta}(1)
    = \llbracket\tilde\Sigma_{r\theta}\rrbracket,
\end{equation}
with the left-hand sides determined entirely by $\psi^{\mi}$ and its
derivatives at $r = 1$, and the jumps prescribed in terms of $\{\zeta,
B(1)\}$ by Eqs.~\eqref{eq:normal_bc}--\eqref{eq:shear_bc}. The interior stream function satisfies the forced Stokes equation; on taking the curl of the momentum balance we obtain
\begin{equation}
\label{eq:biharmonic_stream_function}
    \Delta^2\psi^\mi
        = \alpha \left[\nabla\times\left(\nabla\cdot \tilde{\bQ}\right)\right]
        \cdot\uvc{z},
\end{equation}
which in modal form reads
\begin{equation}
\label{eq:forced_biharmonic}
    \mathcal{S}_m[\psi^\mi]
        = \alpha\left[B''
        + \frac{3}{r} B'
        + \frac{m^2}{r^2} B - \frac{2 \mi m}{r^2} \left(r A\right)'\right].
\end{equation}
For prescribed $\{A(r), B(r), \zeta\}$, this
problem is closed by (a) regularity at the origin, $\psi^\mi(0) =
\psi^{\mi\prime}(0) = 0$; and (b) the two dynamic conditions
\eqref{eq:sigma_star}. The interface then moves with the local flow, and this
kinematic condition yields,
\begin{equation}
\label{eq:kinematic_growth}
    \sigma \zeta = \mi m\, \psi^\mi(1).
\end{equation}
The structure of the stability problem is now apparent. Because the Stokes
problem is quasi-static, the flow is instantaneously slaved to the state of
the system: Eq.~\eqref{eq:forced_biharmonic} is linear, and its boundary
data are linear in $\{A, B, \zeta\}$, so superposition gives
\begin{equation}
\label{eq:psi_solution_operator}
    \psi^\mi(r) = \mathcal{K}^{Q}_m[A, B](r)
        + \zeta\,\mathcal{K}^{\zeta}_m(r),
\end{equation}
where $\mathcal{K}^{Q}_m$ maps the active bulk forcing at zero interfacial
displacement to the flow, and $\mathcal{K}^{\zeta}_m$ is the flow driven by
a unit shape perturbation through its elastic and alignment tractions. The flow is thus determined instantaneously by the state $\{\zeta, A(r), B(r)\}$; to close the problem we require the evolution of these director profiles themselves, to which we turn next.
\subsection{Linearized nematodynamics}

The final step in our linear analysis is the evolution of the perturbations to the nematic tensor. Linearizing the evolution equation~\eqref{eq:dimensionless_nematic_evolution} about the base state
(with $d_R = 0$, and the alignment interaction deferred to the boundary
conditions), we find that the perturbations evolve as
\begin{align}
\label{eq:linA}
\sigma A &= \mathcal{D}_m A - \frac{4\mi m}{r^2} B
+ \mi m\left[\frac{\lambda_+\lambda_-\left(r\psi' - \psi\right)}{r^2}
- 2a\,\psi\right], \\
\label{eq:linB}
\sigma B &= \mathcal{D}_m B + \frac{4\mi m}{r^2} A
- \frac{\lambda_+}{2}\left(\psi'' - \frac{\psi'}{r}\right)
- \frac{m^2 \lambda_-}{2r^2}\,\psi,
\end{align}
where $\mathcal{D}_m \equiv \partial_r^2 + r^{-1}\partial_r - (m^2+4)/r^2$
is the radial Laplacian at azimuthal index $m$; we have also introduced $\lambda_\pm(r) \equiv 1 \pm 2ar^2$ for the sake of notational simplicity. Both equations are supplemented by regularity at the origin.

As highlighted before, in our LSA, the alignment interaction enters not as a bulk term but through the boundary conditions. Linearizing the Robin condition~\eqref{eq:robin} on the perturbed interface and on accounting for the displaced boundary, the rotated tangent and normal~\eqref{eq:linearized_tangent}--\eqref{eq:linearized_normal}, and the perturbed nematic tensor~\eqref{eq:Q_at_interface} yield, at $r = 1$,
\begin{align}
\label{eq:nematic_bc_A}
A'(1) &= 4 a W A(1) + 2a\left(4 a W - 1\right)\zeta, \\
\label{eq:nematic_bc_B}
B'(1) &= 2 a W B(1) + \mi m \left(2 a + W\right)\zeta.
\end{align}
These conditions highlight that a shape fluctuation
$\zeta \neq 0$ rotates the local tangent and displaces the boundary through
the ordered base state, and the anchoring, with strength $W$, rotates the
director field accordingly to follow these shape perturbations. It is through Eqs.~\eqref{eq:nematic_bc_A}--\eqref{eq:nematic_bc_B} alone that the shape forces the nematic field.

The above analysis completes our formulation of the eigenvalue problem. We now highlight the structure of the problem and outline the numerical procedure to solve for the unknown growth rate. The state of the system, or the eigenvector, is the triplet $\{\zeta, A, B\}$; the stream-function $\psi$ is determined quasi-statically by the Stokes problem~\eqref{eq:forced_biharmonic} with the dynamic conditions~\eqref{eq:sigma_star} and regularity at the origin. Discretizing the radial profiles by a second-order accurate finite-difference scheme and appending $\psi$ to the state vector, the linearized system takes the form of a generalized eigenvalue problem,
\begin{equation}
\label{eq:gevp}
\mathcal{L}
\begin{pmatrix} \zeta \\ A \\ B \\ \psi \end{pmatrix}
= \sigma\, \mathcal{B}
\begin{pmatrix} \zeta \\ A \\ B \\ \psi \end{pmatrix},
\qquad
\mathcal{L} =
\begin{pmatrix}
0 & 0 & 0 & \mi m\, \mathrm{e}_1 \\
\Box & \mathcal{D}_m & -4\mi m/r^2 & \mathcal{V}_A \\
\Box & 4\mi m/r^2 & \mathcal{D}_m & \mathcal{V}_B \\
\Box & 0 & 0 & \mathcal{S}_m
\end{pmatrix},
\qquad
\mathcal{B} = \mathrm{diag}(1, \mathbb{I}, \mathbb{I}, 0),
\end{equation}
where, $\mathrm{e}_1$ evaluates $\psi$ at $r = 1$ as part of the kinematic condition, $\mathcal{V}_{A,B}$ are the flow-forcing operators appearing in Eqs.~\eqref{eq:linA}--\eqref{eq:linB}, the fourth row is the forced Stokes problem, and the entries marked as $\Box$ collect the couplings to the shape: $\zeta$ enters the nematodynamic problem through the boundary conditions~\eqref{eq:nematic_bc_A}--\eqref{eq:nematic_bc_B} and the Stokes problem through the elastic and alignment tractions in~\eqref{eq:normal_bc}--\eqref{eq:shear_bc}, imposed as boundary rows of the discretized operators. For each mode $m$ and parameter set $(\alpha, K_B)$, we assemble these matrices numerically and compute the spectrum; the eigenvalue with the largest real part defines the growth rate $\sigma(m)$, and the associated eigenvector provides us with the unstable mode structure $(\zeta, A, B, \psi)$.

\begin{figure}[t!]
    \centering
    \includegraphics[width=0.9\textwidth]{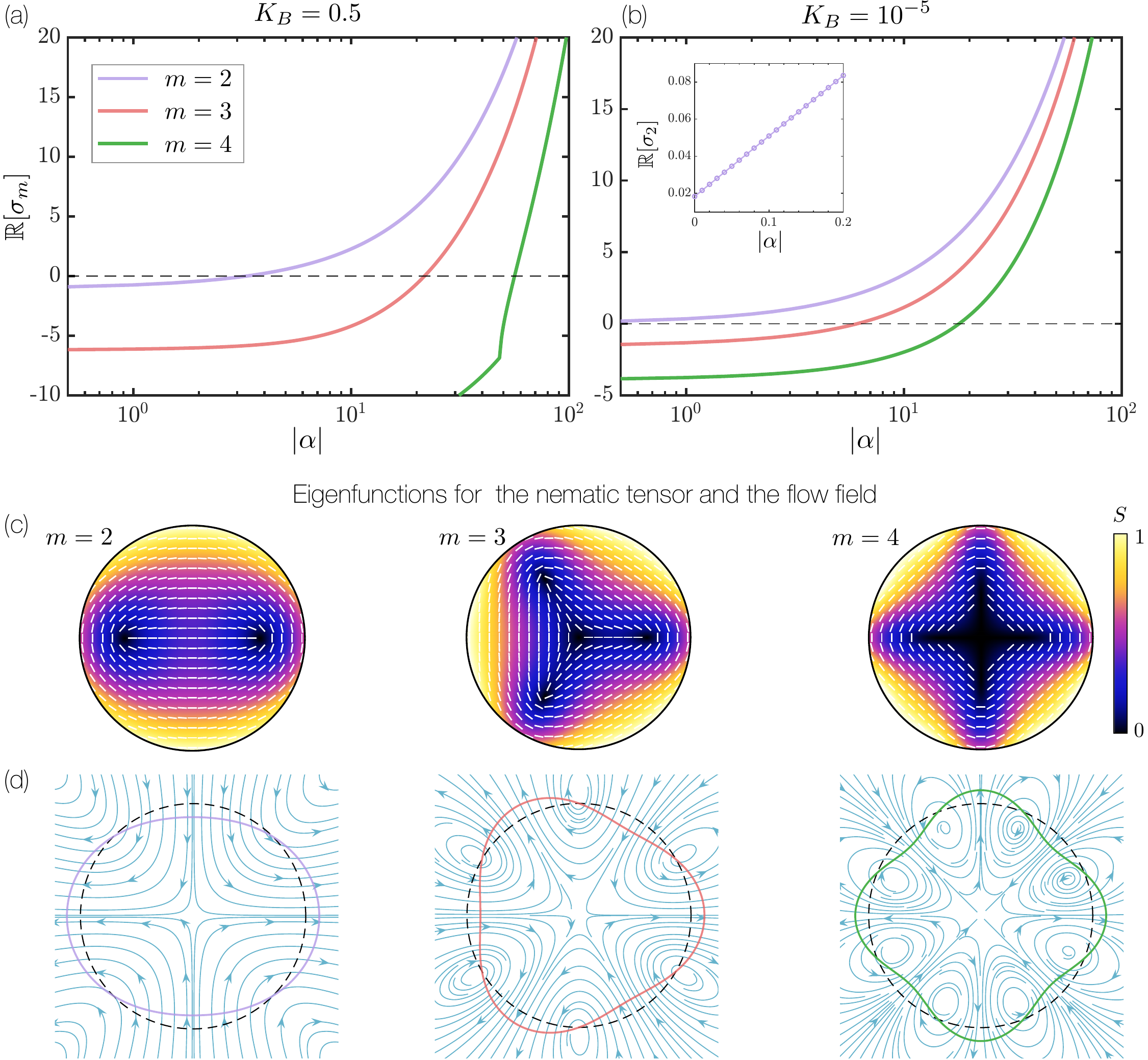}
    \caption{\textbf{Linear stability of the circular vortex state}.
(a,b)~Growth rate $\mathbb{R}[\sigma_m]$ of the interfacial branch for azimuthal modes $m=2,3,4$ versus activity magnitude $|\alpha|$ (extensile, $\alpha<0$), for (a) a stiff ring, $K_B=0.5$, and (b) a floppy ring, $K_B=10^{-5}$, at anchoring strength $\beta=20$. For the stiff ring, the passive circle is stable, and activity destabilizes the modes in order of increasing $m$; for the floppy ring, the structure is unchanged except that $m=2$ is unstable already without activity (inset: $\mathbb{R}[\sigma_2]$ for $|\alpha| \leq0.2 $). (c)~Eigenfunctions of the nematic tensor for $m=2,3,4$: nematic order $S \in [0,1]$ and director (white segments) of $\bQ^{(0)}+\varepsilon\tilde\bQ$ with a visualization amplitude $\varepsilon=0.12$. The order is depleted along an $m$-armed star aligned with the lobes; for $m=2$ the two depleted lobes are the nascent pair of $+1/2$ defects into which the central $+1$ core splits. (d)~Instantaneous streamlines of the perturbation velocity for the same eigenmodes, interior and exterior to the vesicle, showing $2m$ recirculation cells with outflow along the lobes. Parameters: $N_r =120$ radial grid points were used to solve the generalized eigenvalue problem.}
    \label{fig:dispersion}
\end{figure}
\vspace*{-2mm}

\subsection{Results from the stability analysis}
Figure~\ref{fig:dispersion} summarizes the results from linear stability analysis of the circular vortex state as obtained from the solution of the generalized eigenvalue problem. Figure~\ref{fig:dispersion}(a,b) shows the real part of the growth rate $\mathbb{R}[\sigma_m]$ of the interfacial perturbation for the azimuthal modes $m=2, 3, 4$ as a function of activity, for a stiff ring ($K_B=0.5$, panel (a)) and a floppy one ($K_B=10^{-5}$, panel (b)); panels (c,d) show the corresponding eigenfunctions of the nematic tensor and of the flow. For the stiff ring, the passive circular state is linearly stable, and at $\alpha=0$ all shape modes decay at rates that increase roughly as $\sim m^3$ as bending dampens the higher harmonics. Extensile activity destabilizes the modes sequentially, $m=2$ (elliptical mode) first at $|\alpha|=\alpha_c(m{=}2)\approx 3$, followed by $m=3$ (trefoil) and $m=4$ (quatrefoil) at successively larger activities. The eigenvalues of this dominant branch are real, and the onset is stationary as the circle loses stability to a steadily growing $m$-fold deformation. Beyond onset, the growth rates increase with $|\alpha|$, and the dynamics become activity-dominated. The eigenfunctions (in Fig.~\ref{fig:dispersion}(c,d)) illustrate how the instability proceeds. The nematic close to the boundary remains roughly tangential while the interior order is depleted along an $m$-armed star aligned with the lobes of the deformation. The associated perturbation flow is a $2m$-cell recirculation with outflow along the lobes and inflow along the indents as shown by dashed lines in Fig.~\ref{fig:dispersion}(d). The active stress generated by the distorted texture drives precisely the flow that amplifies the distortion. At linear order, this positive feedback competes with bending-driven relaxation of the vesicle, and the threshold $\alpha_c(m)$ marks where the amplification rate first exceeds the relaxation rate, beyond which the mode grows exponentially with $\mathbb{R}[\sigma_m] >0$.

Lowering the bending rigidity preserves this structure (see Fig.~\ref{fig:dispersion}(b)), with the activity thresholds shifted to smaller $|\alpha|$. However, we find one important  qualitative change: the $m=2$ branch is unstable in the absence of activity, with $\mathbb{R}[\sigma_2](\alpha=0)>0$ (inset of Fig.~\ref{fig:dispersion}(b)). This is a passive instability of the vortex state and is driven entirely by the alignment traction arising from the coupling between the suspension and the vesicle boundary via the interfacial force Eq.~\eqref{eq:fibal}. Mechanistically, when the vesicle boundary is perturbed like an ellipse ($m=2$ mode), the suspension adjacent to the boundary reorients to follow the rotated tangent, and the resulting alignment traction is in phase with the deformation as it pushes the lobes outward and pulls the indents inward at a rate the weak bending resistance cannot suppress. The selectivity of this channel is visible in the same panel, as modes $m\geq3$ remain damped at $\alpha=0$, because for these, the traction of the adjusted texture opposes the deformation; only the $m=2$ mode is amplified. This passive instability is the suspension analog of classical liquid-crystal energetics. A nematic liquid crystal confined to a circular cavity with tangential anchoring prefers two half-integer defects over a central $+1$ vortex \cite{de1995physics}, and free nematic droplets elongate into tactoids to reconcile tangential anchoring with a nearly uniform interior, expelling the defects toward the tips \cite{chandler2023, chandler2024}; in either case, the circular vortex state relaxes through elongation and defect rearrangement. Our model carries no such bulk energetic contribution for the suspension of rods. The suspension supports no distortional stress, and the anchoring enters only through the interfacial traction, yet the phenomenology is the same: boundary-driven alignment, elongation, and the fission of the $+1$ core into two $+1/2$ defects. Indeed, although we do not show it here, the instability survives the addition of the elastic energies of liquid crystals. A $\bQ$-tensor liquid-crystal model with elastic (Ericksen) stresses and Rapini--Papoular anchoring \cite{chandler2024} exhibits the same purely elastic $m=2$ instability, with a neutral curve of similar structure. We finally note that our linear analysis ignored rotational diffusion in the suspension; one can stabilize this passive instability at $\alpha = 0$ by having a finite $d_R$.



\section{Results from nonlinear simulations}\label{sec:nonlinear}
We now turn to the nonlinear dynamics of the active vesicle, computed with the volume-conserving spectral immersed boundary method as outlined in Sec.~\ref{sec:numerics}. In our numerical exploration, the stretching modulus $K_S$ is set large enough that the ring is effectively inextensible; with the enclosed area conserved by construction of our numerical scheme, morphology is governed by the competition between the activity $\alpha$ and the bending rigidity $K_B$. Beyond the interior director field with its topological defects, and the instantaneous flow field, we quantify the morphodynamics in two ways. The Fourier amplitudes of the interface,
\begin{equation}
    \label{eq:mode_amplitudes}
    a_m(t) = \left| \frac{1}{2\pi} \oint R(\theta,t)\, e^{-im\theta}\, d\theta \right|,
\end{equation}
measured about the instantaneous vesicle center, resolve the deformation into the azimuthal modes of the linear theory. Complementing this modal decomposition for the interface, we introduce the gyration tensor of the vesicle,
\begin{equation}
    \label{eq:gyration}
    \mathbf{G}(t) = \int_0^1 \left[\bX(s,t) - \bX_c(t)\right]
    \left[\bX(s,t) - \bX_c(t)\right] \md s,
\end{equation}
where $\bX_c(t)$ is the vesicle center of mass. Its dominant eigenvector identifies the instantaneous major axis of the shape, oriented at an angle $\theta(t)$ to the positive $x$ axis, while its eigenvalues $\lambda_1 \geq \lambda_2$ define the asphericity parameter,
\begin{equation}
    \label{eq:asphericity}
    \mathcal{A}(t) =
    \frac{\left[\lambda_1(t) - \lambda_2(t)\right]^2}
         {\left[\lambda_1(t) + \lambda_2(t)\right]^2}.
\end{equation}
The asphericity vanishes for a circle and grows with elongation, providing a single scalar measure of the departure from the circular base state. For weak deformations at leading order, we find that $\mathcal{A} \propto a_2^2$; the asphericity thus serves as a nonlinear generalization of the elliptical mode amplitude, and we use the Fourier amplitudes to monitor the higher harmonics, $m \geq 3$. The linear theory predicts an ordered hierarchy of thresholds, $\alpha_c(m{=}2) < \alpha_c(m{=}3) < \alpha_c(m{=}4)$, at which successive shape modes destabilize; here we show how this cascade of instabilities gives rise to distinct emergent morphologies of the vesicle: static, rotating, fluctuating, and ultimately turbulent states, whose character is revealed by the nonlinear simulations. We now describe these states in order of increasing activity at a fixed bending stiffness. \\

\begin{figure}[H]
    \centering    \includegraphics[width=0.8\linewidth]{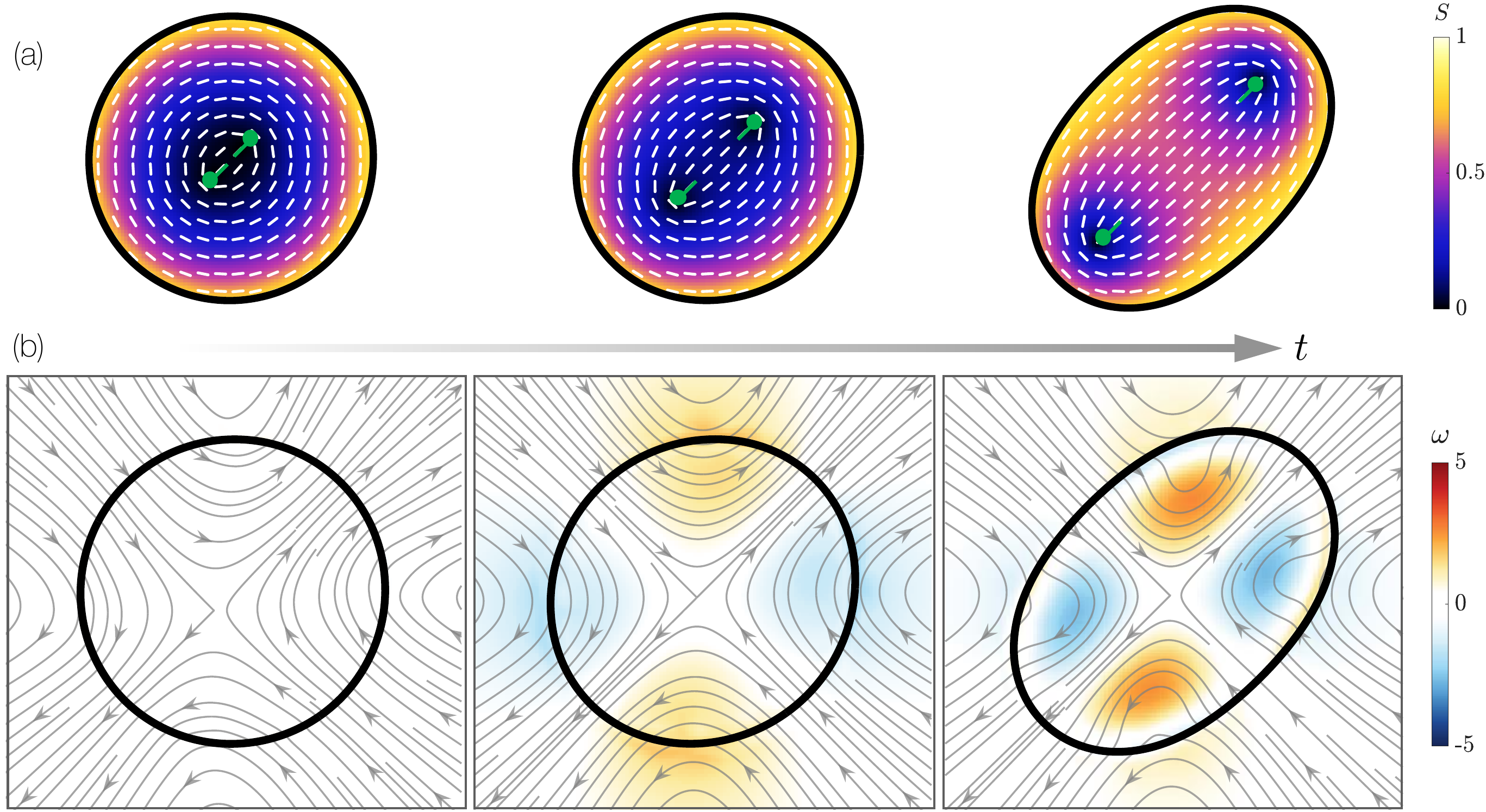}
    \caption{\textbf{Static spindle.} (a)~Time sequence of the vesicle shape and the interior nematic texture beyond the $m=2$ instability threshold, showing the nematic order parameter $S$ (in color) and the director field (white segments). As the vesicle elongates, the central $+1$ vortex core unbinds into two $+1/2$ disclinations (green comet-like markers, with tails indicating defect orientation), which migrate along the major axis and pin at the high-curvature poles, forming a spindle texture. (b)~Corresponding instantaneous streamlines and vorticity field $\omega$ (in color) interior and exterior to the vesicle. The flow develops the four-cell recirculating structure of the $m=2$ eigenmode, with outflow along the lobes (see Supplementary Movie 1). Parameters: $K_B = 5 \times 10^{-5}$, $d_R = 0.5$, $\beta = 20$, and $\alpha = -10$.}
    \label{fig:spindle}
\end{figure}

\noindent  \textbf{Static spindle:} Beyond the first threshold $\alpha_c(m{=}2)$ of the linear theory, the elliptical mode destabilizes while all higher modes remain damped (Fig.~\ref{fig:dispersion}(b)), and the vesicle elongates before saturating into a stationary elliptical shape (Fig.~\ref{fig:spindle}(a)); as is typical of a symmetry-breaking bifurcation, the orientation of the major axis is selected spontaneously by perturbations in our simulations. 
The shape transition is accompanied by a topological rearrangement of the interior fluid, anticipated by the eigenmodes of Fig.~\ref{fig:dispersion}(c): elongation relieves the confinement at the center, and the $+1$ charge at the vortex core unbinds into its two constituent $+1/2$ disclinations. The defects then migrate along the major axis to the high-curvature poles, where the tangential anchoring pins them (Fig.~\ref{fig:spindle}(a)). The resulting texture resembles the spindle configuration, self-organized by the deformation it sustains: the elongated boundary accommodates the defects, and their polar placement in turn stabilizes the elongated shape. The accompanying flow is the $2m$-cell recirculation predicted by the $m=2$ eigenmode, with outflow along the lobes that strengthens as the defects separate (Fig.~\ref{fig:spindle}(b)). At these activities, the active flows are too weak to disrupt this arrangement, and the vesicle reaches a true steady state. \\

\begin{figure}[H]
    \centering
    \includegraphics[width=1\linewidth]{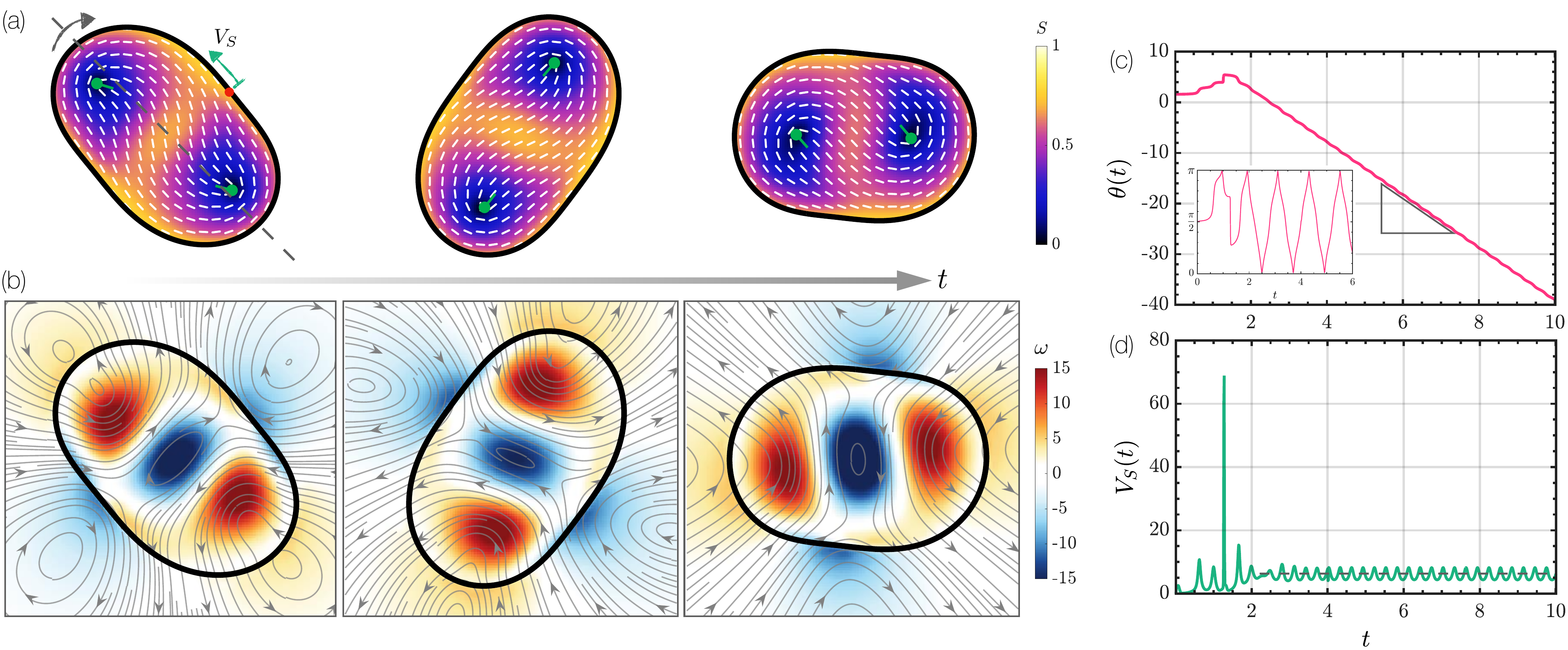}
    \caption{\textbf{Rotor dynamics.} (a)~Time sequence of the rotating vesicle. The spindle texture, with the two $+1/2$ disclinations pinned at the poles, rotates with the shape nearly as a rigid body; the curved arrow indicates the sense of rotation and the dashed line the instantaneous major axis. A tracked material point (red) simultaneously slides along the contour with tank-treading velocity $V_S$ (green arrow), directed against the rotation. (b)~Corresponding streamlines and vorticity: the interior flow organizes into counter-rotating vortices flanking the central region, and the broken mirror symmetry of this arrangement sustains the rotation (see Supplementary Movie 2). (c)~Orientation angle $\theta(t)$ of the major axis, obtained from the dominant eigenvector of the gyration tensor $\mathbf{G}(t)$. After an initial transient, the unwrapped angle decreases linearly in time, defining a constant angular velocity $\dot{\theta}$ (slope indicated); the negative slope corresponds to the clockwise rotation. Inset: the same angle wrapped to $[0,\pi)$, revealing the small periodic wobbles, imparted by the weakly unstable higher harmonics, that ride on the uniform rotation. (d)~Tank-treading velocity $V_S(t)$ extracted from the simulations. Following the transient, $V_S$ fluctuates weakly above its steady-state mean (gray dashed line); its sign, opposite to $\dot{\theta}$, confirms that material points tread against the  rotation. Parameters: $K_B = 5 \times 10^{-5}$, $d_R = 0.5$, $\beta = 20$, and $\alpha = -90$.}
    \label{fig:rotor}
\end{figure}

\noindent \textbf{Rotors:} As activity increases, the stationary spindle gives way to persistent motion. The active flows driven by the stress gradients concentrated at the two defect cores grow stronger, and the mirror-symmetric flow of the static spindle breaks spontaneously. The interior flow loses its reflection symmetry about the major axis and acquires a rotational component reminiscent of dynamics observed previously in active drops \cite{gao2017self, young2021, nejad2023spontaneous}. Through the kinematic boundary condition at the interface, this flow entrains the boundary, and the elongated vesicle rotates nearly as a rigid body (see Fig.~\ref{fig:rotor}(a)): a dynamical phase which we call a \textit{rotor}. Close to the onset of this phase, the symmetry-broken circulation develops only slowly; the vesicle exhibits long transients of irregular shape fluctuations before settling into steady rotation. As activity is progressively increased,  the steady rotation is established rapidly.

We highlight that at these activities, the linear theory already predicts several unstable shape modes, with $m = 3, 4$ having crossed their thresholds alongside $m = 2$; the rotor is thus not a pure consequence of the elliptical instability alone. The dynamical content of the state is nonetheless dominated by $m = 2$: the shape remains essentially elliptical, and the unstable higher harmonics survive only as weak, unsteady perturbations on the rotating ellipse. To quantify the motion, we track the orientation $\theta(t)$ of the major axis, obtained from the dominant eigenvector of the gyration tensor. As shown in Fig.~\ref{fig:rotor}(c), the unwrapped angle evolves linearly in the steady state, defining a constant angular velocity; the angle wrapped to $[0,\pi)$ (inset) resolves, superimposed on this uniform rotation, the small periodic wobbles imparted by the higher harmonics. The vesicle in the figure rotates clockwise, but this sense is selected by initial conditions rather than by any imposed torque or asymmetry; the dynamics has no preferred chirality.
Rigid rotation of the shape, however, does not exhaust the boundary kinematics. Tracking material points along the contour reveals that they additionally slide along the boundary, performing a tank-treading motion classically observed for vesicles and filaments in simple shear flow \cite{kantsler2005, liu2018morphological}. Notably, the tank-treading proceeds against the rotation of the shape: for the clockwise rotor of Fig.~\ref{fig:rotor}, material points tread counterclockwise along the contour. To quantify this, we note that the large stretching modulus keeps the ring effectively inextensible, $\bX_s \cdot \bX_s = 1$, so that $\partial_t \bX_s \cdot \bX_s = 0$. We then posit a simplified Ansatz for the boundary velocity \cite{liu2018morphological}:
\begin{equation}
    \partial_t \bX(s) = \boldsymbol{\Omega} \times \left[\bX(s) -
    \bX_c\right] + V_S\, \bX_s.
\end{equation}
Here we have decomposed the motion into a rigid rotation with angular velocity $\boldsymbol{\Omega} = \dot{\theta}\,\uvc{z}$ and a tangential slip of material points along the contour with tank-treading velocity $V_S$; inextensibility requires $V_S$ to be constant along the boundary. Using the measured $\dot{\theta}$, we extract an arc-length averaged $V_S(t)$ from the simulations and find it approximately constant in the steady rotor state, with a sign opposite to that of $\dot{\theta}$ (Fig.~\ref{fig:rotor}(d)), confirming the counter-rotating tread. \\

\begin{figure}[H]
    \centering
    \includegraphics[width=1\linewidth]{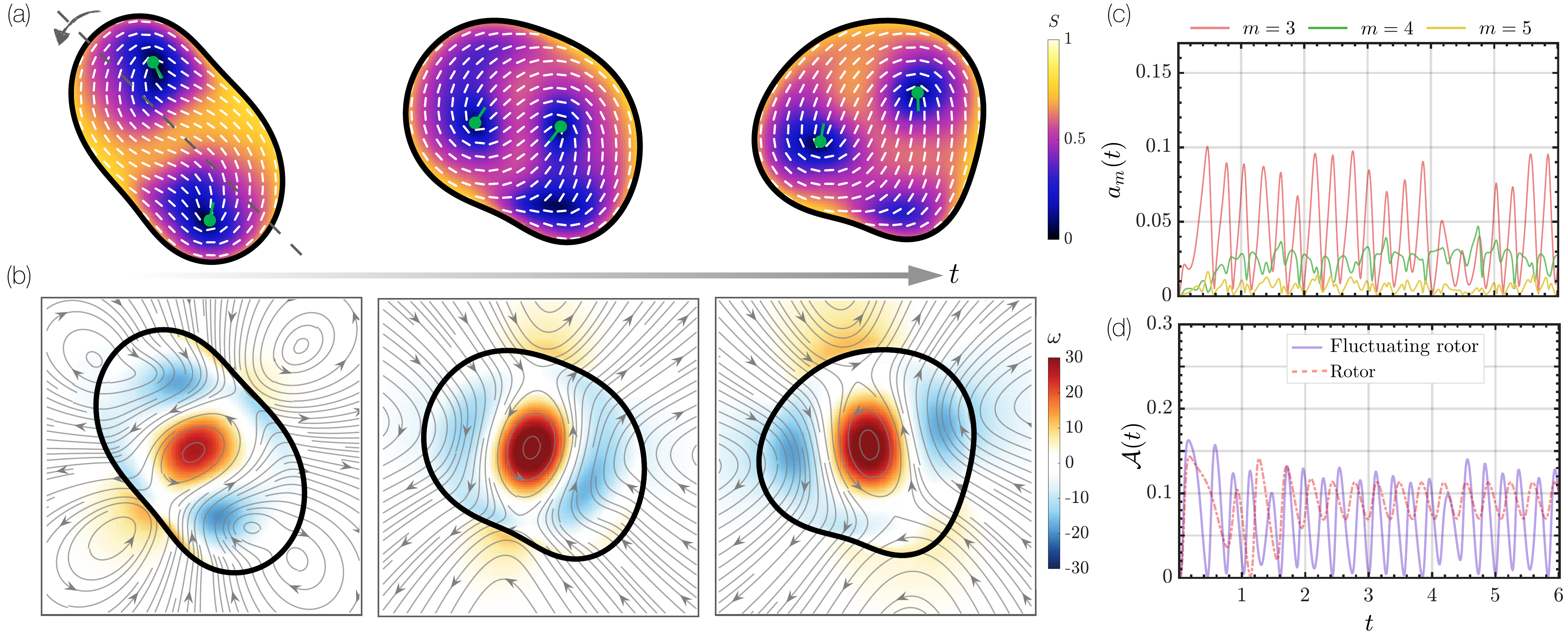}
    \caption{\textbf{Fluctuating rotor.} (a)~Time sequence of the vesicle shape and nematic texture. The two $+1/2$ disclinations are no longer pinned at the poles and instead orbit one another in the interior, while the boundary wobbles as it spins, with unsteady trefoil deformations superimposed on the rotating elliptical shape. This realization rotates counterclockwise, in contrast to the clockwise rotor of Fig.~\ref{fig:rotor}, underscoring that the sense of rotation is selected by initial conditions and the dynamics has no preferred chirality. (b)~Corresponding streamlines and vorticity. The circulating defects organize a strong central vortex that sustains the global rotation, but the surrounding flow is visibly more disordered than in the rotor state (see Supplementary Movie 3). (c)~Shape-mode amplitudes $a_m(t)$: in contrast to the rotor, the $m=3$ amplitude fluctuates persistently about a finite value, the $m=4$ mode contributes weaker modal fluctuations of its own, while the $m=5$ content remains negligible. (d)~The asphericity parameter $\mathcal{A}(t)$ further captures the unsteadiness of the shape, where the rotor exhibits only weak, regular oscillations about its mean, while the fluctuating rotor displays stronger and irregular oscillations. Parameters: $K_B = 5 \times 10^{-5}$, $d_R = 0.5$, $\beta = 20$, and $\alpha = -100$.}
    \label{fig:fluctuating_rotor}
\end{figure}

\noindent \textbf{Fluctuating rotor:} The dominance of the elliptical mode in the rotor phase rests on the separation of the linear growth rates, $\sigma_3 < \sigma_2$. As activity increases, this gap closes, and once the two branches become comparable, the trefoil deformation can no longer be shed, and the rotor loses its steady shape. The interior flows are now strong enough to unpin the two $+1/2$ defects from the elliptical poles, and the unbound defects orbit one another in the interior (see Fig.~\ref{fig:fluctuating_rotor}(a)). The accompanying flow retains a net circulation but is visibly more disordered than in the rotor state as seen in Fig.~\ref{fig:fluctuating_rotor}(b). As localized concentrations of active stress, the circulating defects continually force the nearby membrane, and this forcing sustains the trefoil deformation. We term this resulting morphodynamics a \textit{fluctuating rotor}: the global circulation and the associated rotation persist, but the boundary wobbles as it spins, with unsteady $m=3$ deformations superimposed on the rotating elliptical base shape. 
The modal dynamics reflect this directly; in contrast to the rotor state, the $m=3$ amplitude saturates at a pronounced finite value and fluctuates about it (see Fig.~\ref{fig:fluctuating_rotor}(c)). At these activities, several higher harmonics of shape deformations are also unstable. As seen from Fig.~\ref{fig:fluctuating_rotor}(c), the $m=4$ mode contributes weaker modal fluctuations of its own, while the $m=5$ content remains negligible. The unsteadiness of the shape is further captured by the asphericity $\mathcal{A}(t)$, which we compare directly against the rotor in Fig.~\ref{fig:fluctuating_rotor}(d): the rotor exhibits only weak, regular oscillations of $\mathcal{A}$ about its mean, imparted by the higher harmonics riding on the rotating ellipse, whereas in the fluctuating rotor the oscillations are markedly stronger and irregular, with the shape repeatedly approaching and departing from near-circular conformations as it spins. \\

\begin{figure}[t]
    \centering
    \includegraphics[width=1\linewidth]{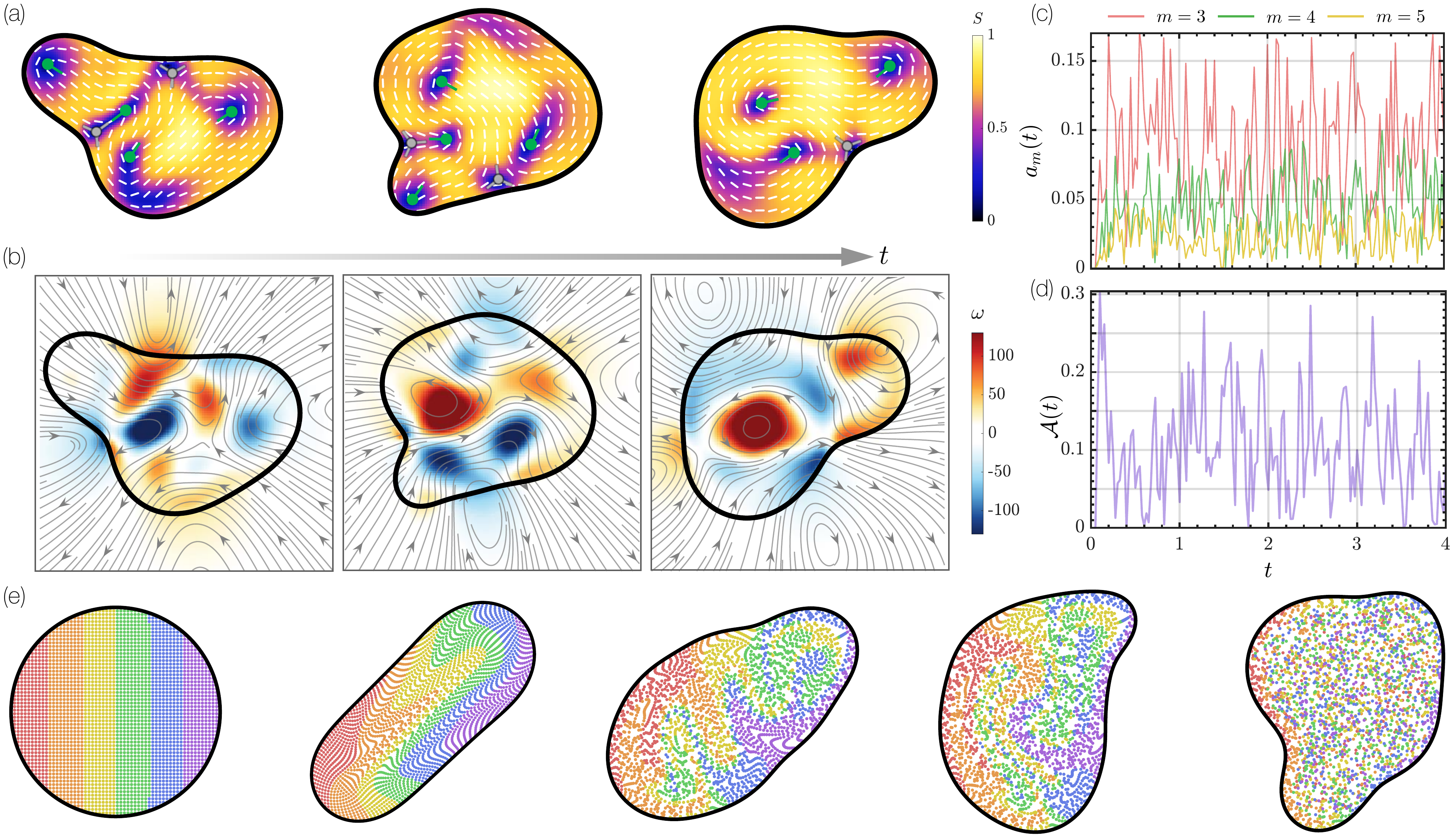}
    \caption{\textbf{Turbulent state.} (a)~Time sequence of the vesicle shape and nematic texture. The interior is populated by multiple defect pairs: $+1/2$ disclinations (green) and $-1/2$ disclinations (gray three-armed stars) continuously nucleate and annihilate, with nucleation occurring predominantly at the boundary. Throughout, the total topological charge is conserved at $+1$, as required by the anchoring on the closed boundary. (b)~Corresponding streamlines and vorticity $\omega$. The interior flow is spatiotemporally chaotic, organized into transient vortices that form and dissolve with the defects; no coherent circulation survives, and the vesicle tumbles irregularly (see Supplementary Movie 4). (c)~Shape-mode amplitudes $a_m(t)$: the amplitudes $a_3$, $a_4$, and $a_5$ all fluctuate strongly about finite values, with no mode settling to a steady amplitude. (d)~ The asphericity parameter $\mathcal{A}(t)$ fluctuates persistently around a mean far from zero, showing large-amplitude departures from the circular state. (e)~Mixing of passive tracers in the interior: a dye field initialized as vertical stripes of distinct colors is rapidly stretched and folded by the chaotic flow, homogenizing the interior and demonstrating the efficient stirring sustained within the vesicle (see Supplementary Movie 5). Parameters: $K_B = 5 \times 10^{-5}$, $d_R = 0.5$, $\beta = 20$, and $\alpha = -300$.}
    \label{fig:turbulence}
\end{figure}
\noindent \textbf{Turbulent state:} At even higher activity, the confined suspension transitions to a state reminiscent of active turbulence. This regime is characterized by the continuous nucleation and annihilation of $\pm 1/2$ defect pairs, which move along chaotic trajectories rather than settling at the boundary or on periodic orbits (see Fig.~\ref{fig:turbulence}(a)). Notably, the defect pairs typically nucleate at the boundary: although the anchoring penalizes departures from tangential order, the active stresses are now strong enough to locally melt the order at the interface, and the boundary layer becomes the dominant source of new defects. The interior flow is spatiotemporally chaotic, organized into transient vortices of alternating sign that form and dissolve with the defects (see Fig.~\ref{fig:turbulence}(b)). This chaotic flow destroys the coherent circulation that drove the rotor states, and the sustained global rotation observed previously ceases. The vesicle instead undergoes irregular tumbling and strong, unsteady shape fluctuations. The linear theory again anticipates the structural signature of this regime: at these activities many shape modes are unstable simultaneously, and the interface responds broadly across them. The amplitudes $a_3$, $a_4$, and $a_5$ now all fluctuate irregularly about finite values, with no separation between the modes of the kind that organized the rotor states (see Fig.~\ref{fig:turbulence}(c)). The asphericity fluctuates strongly and erratically, repeatedly swinging between near-circular conformations and large elongations (see Fig.~\ref{fig:turbulence}(d)); the interface acts as a mechanical readout of the interior turbulence, continually deforming into transient higher-order modes. Finally, we find that the chaotic interior flow efficiently stirs the enclosed fluid: passive tracers, initialized in stripes of distinct colors, are stretched and folded by the flow and rapidly homogenize throughout the vesicle (see Fig.~\ref{fig:turbulence}(e) and Supplementary Movie 5), highlighting that the turbulent vesicle functions as a self-stirring compartment. \\

\begin{figure}[t]
    \centering
    \includegraphics[width=1\linewidth]{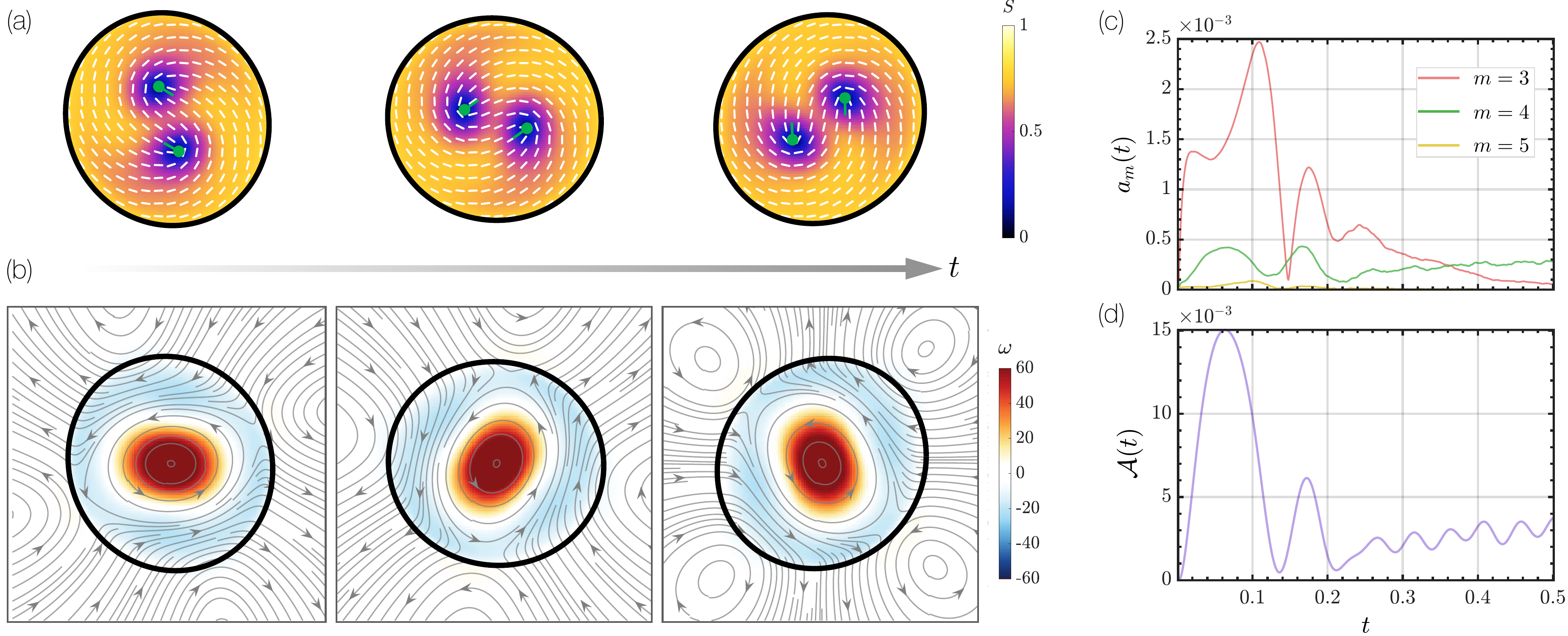}
    \caption{\textbf{Circulating yin-yang state of a stiff vesicle.}
    (a)~Time sequence of the nematic texture inside a vesicle of large bending rigidity. The two $+1/2$ disclinations, with no high-curvature poles to pin them, circulate perpetually on a symmetric orbit about the center in a \textit{yin-yang} configuration. (b)~Corresponding streamlines and vorticity $\omega$: the orbiting pair drives a steady global rotation of the interior fluid, organized around a strong central vortex, while the boundary remains essentially stationary. (c)~Shape-mode amplitudes $a_m(t)$ and (d)~asphericity $\mathcal{A}(t)$, both of which remain small at all times, apart from weak residual oscillations that track the orbiting defect pair: the stiff membrane admits no pronounced deformation, and the interior instability proceeds without any significant morphodynamics. Parameters: $K_B = 50$, $d_R = 0.5$, $\beta = 20$, and $\alpha = -150$.}
    \label{fig:yinyang}
\end{figure}

\noindent \textbf{Circulating yin-yang state:} To isolate the role of boundary deformability, we finally consider vesicles with large bending stiffness. The activity thresholds of the shape modes grow with $K_B$, so a stiff ring is never rigid in the strict sense; sufficiently strong activity will always deform it, but for stiff enough membranes these thresholds lie well beyond the activities considered here; the vesicle remains effectively a circular cavity. Only two states then remain. At low activity, the system reproduces the static circular base state. At high activity, the interior still destabilizes, since the bulk instability of the confined suspension does not require a deformable boundary, and the central $+1$ charge unbinds into two $+1/2$ defects (see Fig.~\ref{fig:yinyang}(a)). With the boundary held nearly circular, however, there are no high-curvature poles to pin the defects, and no static arrangement exists. Instead, the self-generated flows sweep the pair onto a stable, symmetric orbit about the center: the defects circulate perpetually in a \textit{yin-yang} configuration, driving a steady global rotation of the interior fluid within an essentially nondeforming boundary (see Fig.~\ref{fig:yinyang}(b)), as reported previously for active nematics in rigid circular confinement \cite{norton2018}. The shape diagnostics contrast with the quantitative results for deformable vesicles. The higher-mode amplitudes decay after an initial transient and remain two orders of magnitude below their values in the rotor states at comparable activities (see Fig.~\ref{fig:yinyang}(c)): the stiff membrane now damps the modes whose growth organized the morphological cascade. The asphericity likewise remains  $\mathcal{O}(10^{-3})$ (see Fig.~\ref{fig:yinyang}(d)), registering only a weak residual breathing of the boundary driven by the orbiting defect pair.

\section{Conclusion}\label{sec:conclusion}

In this work, we have introduced and analyzed a minimal model of an active vesicle stirred from within and suspended in a passive Newtonian fluid. A closed, nearly inextensible elastic ring, endowed with bending rigidity confines a dilute suspension of extensile active rods and interacts with it through two channels: the flows that the suspension generates and the ring, in deforming, imposes on it, and a Maier--Saupe alignment energy whose action--reaction pair reorients the rods along the membrane and, reciprocally, deforms the membrane to follow the local nematic director. Our modeling permits an exact reduction of the alignment interaction to a Robin boundary condition on the nematic tensor, connecting the penalty-based formulation of our simulations to the weak-anchoring boundary conditions of liquid crystal physics. This reduction yields a closed-form base state, a $+1$ vortex whose tangential order grows quadratically with radius. A linear stability analysis of this base state predicts an ordered hierarchy of activity thresholds, $\alpha_c(m{=}2)<\alpha_c(m{=}3)<\alpha_c(m{=}4)$, at which successive shape modes destabilize and the nonlinear simulations realize these crossings as distinct morphodynamic states. Beyond the first threshold, the vesicle settles into a static spindle with $+1/2$ disclinations pinned at the high-curvature poles. At higher activity, this gives way to a rotor that spins while tank-treading against its own rotation: a kinematic signature classically associated with vesicles in imposed shear, here generated entirely from within. Stronger activity unpins the defects, producing a fluctuating rotor in which the boundary wobbles as it spins, and finally a turbulent state in which boundary-nucleated defect pairs destroy the coherent circulation and efficiently stir the enclosed fluid. Stiff vesicles, by contrast, recover the circulating yin-yang state known from rigid circular confinement~\cite{norton2018, opathalage2019}, identifying boundary deformability, rather than confinement alone, as the ingredient responsible for this sequence. Deformability also opens an instability channel that requires no active stresses: for elastic enclosures with weak bending energy, the elliptical mode is unstable at $\alpha=0$, driven purely by the alignment interaction of the suspension with the boundaries.

Our results suggest several directions for future work. Since each morphological state leaves a hydrodynamic signature in the exterior fluid, the exterior solution decays at large distances as a two-dimensional stresslet whose strength is set by the instantaneous shape. A collection of such active vesicles can therefore interact via these long-range hydrodynamic interactions, giving rise to new emergent dynamics. Even at the level of a pair, the questions remain open: do two rotors synchronize their oscillation phases? Can they attract or repel each other? An analogous question arises in collections. While no individual vesicle is self-propelled, their shape oscillations can phase-lock through the fluid, and such hydrodynamic synchronization is known to generate coherent motion and collective swimming in active matter built from individually non-motile units~\cite{chakrabarti2023collective}. Dense packings of these self-deforming, self-stirring compartments would then be intermediate between an active emulsion and a tissue, inviting comparison with active phase-field and vertex models of confluent cell layers~\cite{kim2021, bi2016}. In the same spirit, the model may be viewed as a minimal model of a cell: a deformable compartment whose shape is selected by the activity it encloses \cite{sciortino2025active}. Making such analogies quantitative, however, will require relaxing some of our simplifying assumptions. Cellular activity is often concentrated in a thin cortical layer rather than distributed through the bulk~\cite{prost2015}; confining the active stress to a boundary layer of tunable thickness would interpolate between the present model and an active cortex. Likewise, adding membrane tension and substrate friction might also allow the symmetry-broken states we find to translate~\cite{tjhung2012}. Finally, the extension to three dimensions, where tangential anchoring on a closed surface fixes the total charge at $+2$, promises correspondingly richer defect dynamics~\cite{keber2014, metselaar2019}.

A second set of questions concerns what the interface fluctuations reveal about the activity they enclose. Although our model is deterministic, in the fluctuating and turbulent states the shape amplitudes $a_m(t)$ fluctuate persistently about finite means, driven entirely by the chaotic active stresses within: the enclosed suspension itself plays the role of a nonequilibrium bath \cite{takatori2020active, villalobos2025active, sciortino2025active}. This suggests a possibility of using the membrane as a probe to infer properties of the confined active matter from the measured shape-fluctuation spectrum, an inverted analogue of flicker spectroscopy of vesicles and red blood cells~\cite{brochard1975, betz2009, turlier2016}. 
Finally, the activity here is spatially uniform by construction, whereas biological activity is often heterogeneous in space. Patterned activity $\alpha(\mathbf{x},t)$, now experimentally accessible through light-activated motor proteins~\cite{zhang2021, lemma2023}, can couple different azimuthal modes which in turn can select higher-order shapes directly, pin or expel disclinations on demand, or convert a rotor into a swimmer. Emergent heterogeneity, through self-propulsion of the constituents poses similar questions. The framework developed here, which involves an exact treatment of the anchoring interaction, a volume-conserving numerical scheme, and exploration of different phases organized by activity and bending rigidity, provides a first step toward these problems, and more broadly toward understanding how soft boundaries sculpt, and are sculpted by, the active flows they confine.

\vspace*{-2.5mm}

\section*{Data Availability Statement}

The in-house code used in this study is available from the corresponding author on reasonable request. 

\vspace*{-2.5mm}

\section*{Author Contributions}
B.C. conceived the project. D.M.M. and B.C. developed the model and performed the stability analysis. D.M.M. developed the numerical framework, performed simulations, and data analysis. D.M.M. and B.C. interpreted the results and wrote the manuscript. The authors declare no conflict of interest.

\vspace*{-2.5mm}

\begin{acknowledgments}
B.C. acknowledges support from the Department of Atomic Energy, Government of India (under project number RTI4019), ANRF ECRG grant 002268, and support from the ICTS CALIBRE grant. The authors thank Anke Lindner, Rituparno Mandal, Samriddhi Sankar Ray, and Michael J. Shelley for various discussions.
\end{acknowledgments}

\bibliographystyle{apsrev4-2}  
\bibliography{refs_brato}

\begin{thebibliography}{64}%
\makeatletter
\providecommand \@ifxundefined [1]{%
 \@ifx{#1\undefined}
}%
\providecommand \@ifnum [1]{%
 \ifnum #1\expandafter \@firstoftwo
 \else \expandafter \@secondoftwo
 \fi
}%
\providecommand \@ifx [1]{%
 \ifx #1\expandafter \@firstoftwo
 \else \expandafter \@secondoftwo
 \fi
}%
\providecommand \natexlab [1]{#1}%
\providecommand \enquote  [1]{``#1''}%
\providecommand \bibnamefont  [1]{#1}%
\providecommand \bibfnamefont [1]{#1}%
\providecommand \citenamefont [1]{#1}%
\providecommand \href@noop [0]{\@secondoftwo}%
\providecommand \href [0]{\begingroup \@sanitize@url \@href}%
\providecommand \@href[1]{\@@startlink{#1}\@@href}%
\providecommand \@@href[1]{\endgroup#1\@@endlink}%
\providecommand \@sanitize@url [0]{\catcode `\\12\catcode `\$12\catcode
  `\&12\catcode `\#12\catcode `\^12\catcode `\_12\catcode `\%12\relax}%
\providecommand \@@startlink[1]{}%
\providecommand \@@endlink[0]{}%
\providecommand \url  [0]{\begingroup\@sanitize@url \@url }%
\providecommand \@url [1]{\endgroup\@href {#1}{\urlprefix }}%
\providecommand \urlprefix  [0]{URL }%
\providecommand \Eprint [0]{\href }%
\providecommand \doibase [0]{https://doi.org/}%
\providecommand \selectlanguage [0]{\@gobble}%
\providecommand \bibinfo  [0]{\@secondoftwo}%
\providecommand \bibfield  [0]{\@secondoftwo}%
\providecommand \translation [1]{[#1]}%
\providecommand \BibitemOpen [0]{}%
\providecommand \bibitemStop [0]{}%
\providecommand \bibitemNoStop [0]{.\EOS\space}%
\providecommand \EOS [0]{\spacefactor3000\relax}%
\providecommand \BibitemShut  [1]{\csname bibitem#1\endcsname}%
\let\auto@bib@innerbib\@empty
\bibitem [{\citenamefont {Marchetti}\ \emph {et~al.}(2013)\citenamefont
  {Marchetti}, \citenamefont {Joanny}, \citenamefont {Ramaswamy}, \citenamefont
  {Liverpool}, \citenamefont {Prost}, \citenamefont {Rao},\ and\ \citenamefont
  {Simha}}]{marchetti2013}%
  \BibitemOpen
  \bibfield  {author} {\bibinfo {author} {\bibfnamefont {M.~C.}\ \bibnamefont
  {Marchetti}}, \bibinfo {author} {\bibfnamefont {J.~F.}\ \bibnamefont
  {Joanny}}, \bibinfo {author} {\bibfnamefont {S.}~\bibnamefont {Ramaswamy}},
  \bibinfo {author} {\bibfnamefont {T.~B.}\ \bibnamefont {Liverpool}}, \bibinfo
  {author} {\bibfnamefont {J.}~\bibnamefont {Prost}}, \bibinfo {author}
  {\bibfnamefont {M.}~\bibnamefont {Rao}},\ and\ \bibinfo {author}
  {\bibfnamefont {R.~A.}\ \bibnamefont {Simha}},\ }\href
  {https://doi.org/10.1103/revmodphys.85.1143} {\bibfield  {journal} {\bibinfo
  {journal} {Rev. Mod. Phys.}\ }\textbf {\bibinfo {volume} {85}},\ \bibinfo
  {pages} {1143} (\bibinfo {year} {2013})}\BibitemShut {NoStop}%
\bibitem [{\citenamefont {Doostmohammadi}\ \emph {et~al.}(2018)\citenamefont
  {Doostmohammadi}, \citenamefont {Ign{\'e}s-Mullol}, \citenamefont {Yeomans},\
  and\ \citenamefont {Sagu{\'e}s}}]{doostmohammadi2018}%
  \BibitemOpen
  \bibfield  {author} {\bibinfo {author} {\bibfnamefont {A.}~\bibnamefont
  {Doostmohammadi}}, \bibinfo {author} {\bibfnamefont {J.}~\bibnamefont
  {Ign{\'e}s-Mullol}}, \bibinfo {author} {\bibfnamefont {J.~M.}\ \bibnamefont
  {Yeomans}},\ and\ \bibinfo {author} {\bibfnamefont {F.}~\bibnamefont
  {Sagu{\'e}s}},\ }\href {https://doi.org/10.1038/s41467-018-05666-8}
  {\bibfield  {journal} {\bibinfo  {journal} {Nat. Commun.}\ }\textbf {\bibinfo
  {volume} {9}},\ \bibinfo {pages} {3246} (\bibinfo {year} {2018})}\BibitemShut
  {NoStop}%
\bibitem [{\citenamefont {Mayer}\ \emph {et~al.}(2010)\citenamefont {Mayer},
  \citenamefont {Depken}, \citenamefont {Bois}, \citenamefont {J{\"u}licher},\
  and\ \citenamefont {Grill}}]{mayer2010}%
  \BibitemOpen
  \bibfield  {author} {\bibinfo {author} {\bibfnamefont {M.}~\bibnamefont
  {Mayer}}, \bibinfo {author} {\bibfnamefont {M.}~\bibnamefont {Depken}},
  \bibinfo {author} {\bibfnamefont {J.~S.}\ \bibnamefont {Bois}}, \bibinfo
  {author} {\bibfnamefont {F.}~\bibnamefont {J{\"u}licher}},\ and\ \bibinfo
  {author} {\bibfnamefont {S.~W.}\ \bibnamefont {Grill}},\ }\href
  {https://doi.org/10.1038/nature09376} {\bibfield  {journal} {\bibinfo
  {journal} {Nature}\ }\textbf {\bibinfo {volume} {467}},\ \bibinfo {pages}
  {617} (\bibinfo {year} {2010})}\BibitemShut {NoStop}%
\bibitem [{\citenamefont {Turlier}\ \emph {et~al.}(2014)\citenamefont
  {Turlier}, \citenamefont {Audoly}, \citenamefont {Prost},\ and\ \citenamefont
  {Joanny}}]{turlier2014}%
  \BibitemOpen
  \bibfield  {author} {\bibinfo {author} {\bibfnamefont {H.}~\bibnamefont
  {Turlier}}, \bibinfo {author} {\bibfnamefont {B.}~\bibnamefont {Audoly}},
  \bibinfo {author} {\bibfnamefont {J.}~\bibnamefont {Prost}},\ and\ \bibinfo
  {author} {\bibfnamefont {J.-F.}\ \bibnamefont {Joanny}},\ }\href
  {https://doi.org/10.1016/j.bpj.2013.11.014} {\bibfield  {journal} {\bibinfo
  {journal} {Biophys. J.}\ }\textbf {\bibinfo {volume} {106}},\ \bibinfo
  {pages} {114} (\bibinfo {year} {2014})}\BibitemShut {NoStop}%
\bibitem [{\citenamefont {Rajput}\ \emph {et~al.}(2025)\citenamefont {Rajput},
  \citenamefont {Kumar}, \citenamefont {Nishikawa},\ and\ \citenamefont
  {Kumar}}]{rajput2025active}%
  \BibitemOpen
  \bibfield  {author} {\bibinfo {author} {\bibfnamefont {A.~S.}\ \bibnamefont
  {Rajput}}, \bibinfo {author} {\bibfnamefont {K.}~\bibnamefont {Kumar}},
  \bibinfo {author} {\bibfnamefont {M.}~\bibnamefont {Nishikawa}},\ and\
  \bibinfo {author} {\bibfnamefont {K.~V.}\ \bibnamefont {Kumar}},\ }\href
  {https://doi.org/10.1101/2025.11.07.683232} {\bibfield  {journal} {\bibinfo
  {journal} {bioRxiv}\ ,\ \bibinfo {pages} {2025}} (\bibinfo {year}
  {2025})}\BibitemShut {NoStop}%
\bibitem [{\citenamefont {Nazockdast}\ \emph {et~al.}(2017)\citenamefont
  {Nazockdast}, \citenamefont {Rahimian}, \citenamefont {Needleman},\ and\
  \citenamefont {Shelley}}]{nazockdast2017cytoplasmic}%
  \BibitemOpen
  \bibfield  {author} {\bibinfo {author} {\bibfnamefont {E.}~\bibnamefont
  {Nazockdast}}, \bibinfo {author} {\bibfnamefont {A.}~\bibnamefont
  {Rahimian}}, \bibinfo {author} {\bibfnamefont {D.}~\bibnamefont
  {Needleman}},\ and\ \bibinfo {author} {\bibfnamefont {M.}~\bibnamefont
  {Shelley}},\ }\href {https://doi.org/10.1091/mbc.E16-02-0108} {\bibfield
  {journal} {\bibinfo  {journal} {Molecular biology of the cell}\ }\textbf
  {\bibinfo {volume} {28}},\ \bibinfo {pages} {3261} (\bibinfo {year}
  {2017})}\BibitemShut {NoStop}%
\bibitem [{\citenamefont {Zidovska}\ \emph {et~al.}(2013)\citenamefont
  {Zidovska}, \citenamefont {Weitz},\ and\ \citenamefont
  {Mitchison}}]{zidovska2013}%
  \BibitemOpen
  \bibfield  {author} {\bibinfo {author} {\bibfnamefont {A.}~\bibnamefont
  {Zidovska}}, \bibinfo {author} {\bibfnamefont {D.~A.}\ \bibnamefont
  {Weitz}},\ and\ \bibinfo {author} {\bibfnamefont {T.~J.}\ \bibnamefont
  {Mitchison}},\ }\href {https://doi.org/10.1073/pnas.1220313110} {\bibfield
  {journal} {\bibinfo  {journal} {Proc. Natl. Acad. Sci. U.S.A.}\ }\textbf
  {\bibinfo {volume} {110}},\ \bibinfo {pages} {15555} (\bibinfo {year}
  {2013})}\BibitemShut {NoStop}%
\bibitem [{\citenamefont {Saintillan}\ \emph {et~al.}(2018)\citenamefont
  {Saintillan}, \citenamefont {Shelley},\ and\ \citenamefont
  {Zidovska}}]{saintillan2018extensile}%
  \BibitemOpen
  \bibfield  {author} {\bibinfo {author} {\bibfnamefont {D.}~\bibnamefont
  {Saintillan}}, \bibinfo {author} {\bibfnamefont {M.~J.}\ \bibnamefont
  {Shelley}},\ and\ \bibinfo {author} {\bibfnamefont {A.}~\bibnamefont
  {Zidovska}},\ }\href {https://doi.org/10.1073/pnas.1807073115} {\bibfield
  {journal} {\bibinfo  {journal} {Proceedings of the National Academy of
  Sciences}\ }\textbf {\bibinfo {volume} {115}},\ \bibinfo {pages} {11442}
  (\bibinfo {year} {2018})}\BibitemShut {NoStop}%
\bibitem [{\citenamefont {Weady}\ \emph {et~al.}(2024)\citenamefont {Weady},
  \citenamefont {Stein}, \citenamefont {Zidovska},\ and\ \citenamefont
  {Shelley}}]{weady2024}%
  \BibitemOpen
  \bibfield  {author} {\bibinfo {author} {\bibfnamefont {S.}~\bibnamefont
  {Weady}}, \bibinfo {author} {\bibfnamefont {D.~B.}\ \bibnamefont {Stein}},
  \bibinfo {author} {\bibfnamefont {A.}~\bibnamefont {Zidovska}},\ and\
  \bibinfo {author} {\bibfnamefont {M.~J.}\ \bibnamefont {Shelley}},\ }\href
  {https://doi.org/10.1103/physrevfluids.9.013102} {\bibfield  {journal}
  {\bibinfo  {journal} {Phys. Rev. Fluids}\ }\textbf {\bibinfo {volume} {9}},\
  \bibinfo {pages} {013102} (\bibinfo {year} {2024})}\BibitemShut {NoStop}%
\bibitem [{\citenamefont {Rautu}\ \emph {et~al.}(2026)\citenamefont {Rautu},
  \citenamefont {Zidovska}, \citenamefont {Saintillan},\ and\ \citenamefont
  {Shelley}}]{rautu2026}%
  \BibitemOpen
  \bibfield  {author} {\bibinfo {author} {\bibfnamefont {S.~A.}\ \bibnamefont
  {Rautu}}, \bibinfo {author} {\bibfnamefont {A.}~\bibnamefont {Zidovska}},
  \bibinfo {author} {\bibfnamefont {D.}~\bibnamefont {Saintillan}},\ and\
  \bibinfo {author} {\bibfnamefont {M.~J.}\ \bibnamefont {Shelley}},\ }\href
  {https://doi.org/10.1103/8n8h-7gx6} {\bibfield  {journal} {\bibinfo
  {journal} {Phys. Rev. X}\ }\textbf {\bibinfo {volume} {16}},\ \bibinfo
  {pages} {021009} (\bibinfo {year} {2026})}\BibitemShut {NoStop}%
\bibitem [{\citenamefont {Adkins}\ \emph {et~al.}(2022)\citenamefont {Adkins},
  \citenamefont {Kolvin}, \citenamefont {You}, \citenamefont {Witthaus},
  \citenamefont {Marchetti},\ and\ \citenamefont {Dogic}}]{adkins2022dynamics}%
  \BibitemOpen
  \bibfield  {author} {\bibinfo {author} {\bibfnamefont {R.}~\bibnamefont
  {Adkins}}, \bibinfo {author} {\bibfnamefont {I.}~\bibnamefont {Kolvin}},
  \bibinfo {author} {\bibfnamefont {Z.}~\bibnamefont {You}}, \bibinfo {author}
  {\bibfnamefont {S.}~\bibnamefont {Witthaus}}, \bibinfo {author}
  {\bibfnamefont {M.~C.}\ \bibnamefont {Marchetti}},\ and\ \bibinfo {author}
  {\bibfnamefont {Z.}~\bibnamefont {Dogic}},\ }\href
  {https://doi.org/10.1126/science.abo5423} {\bibfield  {journal} {\bibinfo
  {journal} {Science}\ }\textbf {\bibinfo {volume} {377}},\ \bibinfo {pages}
  {768} (\bibinfo {year} {2022})}\BibitemShut {NoStop}%
\bibitem [{\citenamefont {Gulati}\ \emph {et~al.}(2024)\citenamefont {Gulati},
  \citenamefont {Caballero}, \citenamefont {Kolvin}, \citenamefont {You},\ and\
  \citenamefont {Marchetti}}]{gulati2024traveling}%
  \BibitemOpen
  \bibfield  {author} {\bibinfo {author} {\bibfnamefont {P.}~\bibnamefont
  {Gulati}}, \bibinfo {author} {\bibfnamefont {F.}~\bibnamefont {Caballero}},
  \bibinfo {author} {\bibfnamefont {I.}~\bibnamefont {Kolvin}}, \bibinfo
  {author} {\bibfnamefont {Z.}~\bibnamefont {You}},\ and\ \bibinfo {author}
  {\bibfnamefont {M.~C.}\ \bibnamefont {Marchetti}},\ }\href
  {https://doi.org/10.1039/D4SM00822G} {\bibfield  {journal} {\bibinfo
  {journal} {Soft Matter}\ }\textbf {\bibinfo {volume} {20}},\ \bibinfo {pages}
  {7703} (\bibinfo {year} {2024})}\BibitemShut {NoStop}%
\bibitem [{\citenamefont {Sciortino}\ \emph {et~al.}(2025)\citenamefont
  {Sciortino}, \citenamefont {Faizi}, \citenamefont {Fedosov}, \citenamefont
  {Frechette}, \citenamefont {Vlahovska}, \citenamefont {Gompper},\ and\
  \citenamefont {Bausch}}]{sciortino2025active}%
  \BibitemOpen
  \bibfield  {author} {\bibinfo {author} {\bibfnamefont {A.}~\bibnamefont
  {Sciortino}}, \bibinfo {author} {\bibfnamefont {H.~A.}\ \bibnamefont
  {Faizi}}, \bibinfo {author} {\bibfnamefont {D.~A.}\ \bibnamefont {Fedosov}},
  \bibinfo {author} {\bibfnamefont {L.}~\bibnamefont {Frechette}}, \bibinfo
  {author} {\bibfnamefont {P.~M.}\ \bibnamefont {Vlahovska}}, \bibinfo {author}
  {\bibfnamefont {G.}~\bibnamefont {Gompper}},\ and\ \bibinfo {author}
  {\bibfnamefont {A.~R.}\ \bibnamefont {Bausch}},\ }\href
  {https://doi.org/10.1038/s41567-025-02839-3} {\bibfield  {journal} {\bibinfo
  {journal} {Nature Physics}\ }\textbf {\bibinfo {volume} {21}},\ \bibinfo
  {pages} {799} (\bibinfo {year} {2025})}\BibitemShut {NoStop}%
\bibitem [{\citenamefont {Kato}\ \emph {et~al.}(2026)\citenamefont {Kato},
  \citenamefont {Br{\'e}mont}, \citenamefont {Hart{\'e}}, \citenamefont
  {Rampnoux}, \citenamefont {Joanny},\ and\ \citenamefont
  {Kellay}}]{kato2026active}%
  \BibitemOpen
  \bibfield  {author} {\bibinfo {author} {\bibfnamefont {A.~N.}\ \bibnamefont
  {Kato}}, \bibinfo {author} {\bibfnamefont {J.}~\bibnamefont {Br{\'e}mont}},
  \bibinfo {author} {\bibfnamefont {E.}~\bibnamefont {Hart{\'e}}}, \bibinfo
  {author} {\bibfnamefont {J.-M.}\ \bibnamefont {Rampnoux}}, \bibinfo {author}
  {\bibfnamefont {J.-F.}\ \bibnamefont {Joanny}},\ and\ \bibinfo {author}
  {\bibfnamefont {H.}~\bibnamefont {Kellay}},\ }\bibfield  {journal} {\bibinfo
  {journal} {arXiv preprint arXiv:2609.21963}\ }\href
  {https://doi.org/10.48550/arXiv.2609.21963} {10.48550/arXiv.2609.21963}
  (\bibinfo {year} {2026})\BibitemShut {NoStop}%
\bibitem [{\citenamefont {Kikuchi}\ \emph {et~al.}(2009)\citenamefont
  {Kikuchi}, \citenamefont {Ehrlicher}, \citenamefont {Koch}, \citenamefont
  {K{\"a}s}, \citenamefont {Ramaswamy},\ and\ \citenamefont
  {Rao}}]{kikuchi2009buckling}%
  \BibitemOpen
  \bibfield  {author} {\bibinfo {author} {\bibfnamefont {N.}~\bibnamefont
  {Kikuchi}}, \bibinfo {author} {\bibfnamefont {A.}~\bibnamefont {Ehrlicher}},
  \bibinfo {author} {\bibfnamefont {D.}~\bibnamefont {Koch}}, \bibinfo {author}
  {\bibfnamefont {J.~A.}\ \bibnamefont {K{\"a}s}}, \bibinfo {author}
  {\bibfnamefont {S.}~\bibnamefont {Ramaswamy}},\ and\ \bibinfo {author}
  {\bibfnamefont {M.}~\bibnamefont {Rao}},\ }\href
  {https://doi.org/10.1073/pnas.0900451106} {\bibfield  {journal} {\bibinfo
  {journal} {Proceedings of the National Academy of Sciences}\ }\textbf
  {\bibinfo {volume} {106}},\ \bibinfo {pages} {19776} (\bibinfo {year}
  {2009})}\BibitemShut {NoStop}%
\bibitem [{\citenamefont {Valei}\ and\ \citenamefont
  {Shendruk}(2025)}]{valei2025scaling}%
  \BibitemOpen
  \bibfield  {author} {\bibinfo {author} {\bibfnamefont {Z.~K.}\ \bibnamefont
  {Valei}}\ and\ \bibinfo {author} {\bibfnamefont {T.~N.}\ \bibnamefont
  {Shendruk}},\ }\href {https://doi.org/10.1103/w3gp-knnp} {\bibfield
  {journal} {\bibinfo  {journal} {Physical Review Letters}\ }\textbf {\bibinfo
  {volume} {135}},\ \bibinfo {pages} {138301} (\bibinfo {year}
  {2025})}\BibitemShut {NoStop}%
\bibitem [{\citenamefont {Callan-Jones}\ \emph {et~al.}(2008)\citenamefont
  {Callan-Jones}, \citenamefont {Joanny},\ and\ \citenamefont
  {Prost}}]{callanjones2008}%
  \BibitemOpen
  \bibfield  {author} {\bibinfo {author} {\bibfnamefont {A.~C.}\ \bibnamefont
  {Callan-Jones}}, \bibinfo {author} {\bibfnamefont {J.-F.}\ \bibnamefont
  {Joanny}},\ and\ \bibinfo {author} {\bibfnamefont {J.}~\bibnamefont
  {Prost}},\ }\href {https://doi.org/10.1103/physrevlett.100.258106} {\bibfield
   {journal} {\bibinfo  {journal} {Phys. Rev. Lett.}\ }\textbf {\bibinfo
  {volume} {100}},\ \bibinfo {pages} {258106} (\bibinfo {year}
  {2008})}\BibitemShut {NoStop}%
\bibitem [{\citenamefont {Hawkins}\ \emph {et~al.}(2011)\citenamefont
  {Hawkins}, \citenamefont {Poincloux}, \citenamefont {B{\'e}nichou},
  \citenamefont {Piel}, \citenamefont {Chavrier},\ and\ \citenamefont
  {Voituriez}}]{hawkins2011spontaneous}%
  \BibitemOpen
  \bibfield  {author} {\bibinfo {author} {\bibfnamefont {R.~J.}\ \bibnamefont
  {Hawkins}}, \bibinfo {author} {\bibfnamefont {R.}~\bibnamefont {Poincloux}},
  \bibinfo {author} {\bibfnamefont {O.}~\bibnamefont {B{\'e}nichou}}, \bibinfo
  {author} {\bibfnamefont {M.}~\bibnamefont {Piel}}, \bibinfo {author}
  {\bibfnamefont {P.}~\bibnamefont {Chavrier}},\ and\ \bibinfo {author}
  {\bibfnamefont {R.}~\bibnamefont {Voituriez}},\ }\href
  {https://doi.org/10.1016/j.bpj.2011.07.038} {\bibfield  {journal} {\bibinfo
  {journal} {Biophysical journal}\ }\textbf {\bibinfo {volume} {101}},\
  \bibinfo {pages} {1041} (\bibinfo {year} {2011})}\BibitemShut {NoStop}%
\bibitem [{\citenamefont {Tjhung}\ \emph {et~al.}(2012)\citenamefont {Tjhung},
  \citenamefont {Marenduzzo},\ and\ \citenamefont {Cates}}]{tjhung2012}%
  \BibitemOpen
  \bibfield  {author} {\bibinfo {author} {\bibfnamefont {E.}~\bibnamefont
  {Tjhung}}, \bibinfo {author} {\bibfnamefont {D.}~\bibnamefont {Marenduzzo}},\
  and\ \bibinfo {author} {\bibfnamefont {M.~E.}\ \bibnamefont {Cates}},\ }\href
  {https://doi.org/10.1073/pnas.1200843109} {\bibfield  {journal} {\bibinfo
  {journal} {Proc. Natl. Acad. Sci. U.S.A.}\ }\textbf {\bibinfo {volume}
  {109}},\ \bibinfo {pages} {12381} (\bibinfo {year} {2012})}\BibitemShut
  {NoStop}%
\bibitem [{\citenamefont {Gao}\ and\ \citenamefont {Li}(2017)}]{gao2017self}%
  \BibitemOpen
  \bibfield  {author} {\bibinfo {author} {\bibfnamefont {T.}~\bibnamefont
  {Gao}}\ and\ \bibinfo {author} {\bibfnamefont {Z.}~\bibnamefont {Li}},\
  }\href {https://doi.org/10.1103/PhysRevLett.119.108002} {\bibfield  {journal}
  {\bibinfo  {journal} {Physical Review Letters}\ }\textbf {\bibinfo {volume}
  {119}},\ \bibinfo {pages} {108002} (\bibinfo {year} {2017})}\BibitemShut
  {NoStop}%
\bibitem [{\citenamefont {Young}\ \emph {et~al.}(2021)\citenamefont {Young},
  \citenamefont {Shelley},\ and\ \citenamefont {Stein}}]{young2021}%
  \BibitemOpen
  \bibfield  {author} {\bibinfo {author} {\bibfnamefont {Y.~N.}\ \bibnamefont
  {Young}}, \bibinfo {author} {\bibfnamefont {M.~J.}\ \bibnamefont {Shelley}},\
  and\ \bibinfo {author} {\bibfnamefont {D.~B.}\ \bibnamefont {Stein}},\ }\href
  {https://doi.org/10.3934/mbe.2021145} {\bibfield  {journal} {\bibinfo
  {journal} {Math. Biosci. Eng.}\ }\textbf {\bibinfo {volume} {18}},\ \bibinfo
  {pages} {2849} (\bibinfo {year} {2021})}\BibitemShut {NoStop}%
\bibitem [{\citenamefont {Nejad}\ and\ \citenamefont
  {Yeomans}(2023)}]{nejad2023spontaneous}%
  \BibitemOpen
  \bibfield  {author} {\bibinfo {author} {\bibfnamefont {M.~R.}\ \bibnamefont
  {Nejad}}\ and\ \bibinfo {author} {\bibfnamefont {J.~M.}\ \bibnamefont
  {Yeomans}},\ }\href {https://doi.org/10.1103/prxlife.1.023008} {\bibfield
  {journal} {\bibinfo  {journal} {PRX Life}\ }\textbf {\bibinfo {volume} {1}},\
  \bibinfo {pages} {023008} (\bibinfo {year} {2023})}\BibitemShut {NoStop}%
\bibitem [{\citenamefont {Sessa}\ \emph {et~al.}(2026)\citenamefont {Sessa},
  \citenamefont {Cao}, \citenamefont {Pelcovits}, \citenamefont {Powers},\ and\
  \citenamefont {Duclos}}]{sessa2026interfacial}%
  \BibitemOpen
  \bibfield  {author} {\bibinfo {author} {\bibfnamefont {B.~C.}\ \bibnamefont
  {Sessa}}, \bibinfo {author} {\bibfnamefont {F.}~\bibnamefont {Cao}}, \bibinfo
  {author} {\bibfnamefont {R.~A.}\ \bibnamefont {Pelcovits}}, \bibinfo {author}
  {\bibfnamefont {T.~R.}\ \bibnamefont {Powers}},\ and\ \bibinfo {author}
  {\bibfnamefont {G.}~\bibnamefont {Duclos}},\ }\href
  {https://doi.org/10.1103/ytwv-7swk} {\bibfield  {journal} {\bibinfo
  {journal} {Physical Review Research}\ }\textbf {\bibinfo {volume} {8}},\
  \bibinfo {pages} {013120} (\bibinfo {year} {2026})}\BibitemShut {NoStop}%
\bibitem [{\citenamefont {Giomi}\ and\ \citenamefont
  {DeSimone}(2014)}]{giomi2014}%
  \BibitemOpen
  \bibfield  {author} {\bibinfo {author} {\bibfnamefont {L.}~\bibnamefont
  {Giomi}}\ and\ \bibinfo {author} {\bibfnamefont {A.}~\bibnamefont
  {DeSimone}},\ }\href {https://doi.org/10.1103/physrevlett.112.147802}
  {\bibfield  {journal} {\bibinfo  {journal} {Phys. Rev. Lett.}\ }\textbf
  {\bibinfo {volume} {112}},\ \bibinfo {pages} {147802} (\bibinfo {year}
  {2014})}\BibitemShut {NoStop}%
\bibitem [{\citenamefont {Weirich}\ \emph {et~al.}(2019)\citenamefont
  {Weirich}, \citenamefont {Dasbiswas}, \citenamefont {Witten}, \citenamefont
  {Vaikuntanathan},\ and\ \citenamefont {Gardel}}]{weirich2019}%
  \BibitemOpen
  \bibfield  {author} {\bibinfo {author} {\bibfnamefont {K.~L.}\ \bibnamefont
  {Weirich}}, \bibinfo {author} {\bibfnamefont {K.}~\bibnamefont {Dasbiswas}},
  \bibinfo {author} {\bibfnamefont {T.~A.}\ \bibnamefont {Witten}}, \bibinfo
  {author} {\bibfnamefont {S.}~\bibnamefont {Vaikuntanathan}},\ and\ \bibinfo
  {author} {\bibfnamefont {M.~L.}\ \bibnamefont {Gardel}},\ }\href
  {https://doi.org/10.1073/pnas.1814854116} {\bibfield  {journal} {\bibinfo
  {journal} {Proc. Natl. Acad. Sci. U.S.A.}\ }\textbf {\bibinfo {volume}
  {116}},\ \bibinfo {pages} {11125} (\bibinfo {year} {2019})}\BibitemShut
  {NoStop}%
\bibitem [{\citenamefont {Keber}\ \emph {et~al.}(2014)\citenamefont {Keber},
  \citenamefont {Loiseau}, \citenamefont {Sanchez}, \citenamefont {DeCamp},
  \citenamefont {Giomi}, \citenamefont {Bowick}, \citenamefont {Marchetti},
  \citenamefont {Dogic},\ and\ \citenamefont {Bausch}}]{keber2014}%
  \BibitemOpen
  \bibfield  {author} {\bibinfo {author} {\bibfnamefont {F.~C.}\ \bibnamefont
  {Keber}}, \bibinfo {author} {\bibfnamefont {E.}~\bibnamefont {Loiseau}},
  \bibinfo {author} {\bibfnamefont {T.}~\bibnamefont {Sanchez}}, \bibinfo
  {author} {\bibfnamefont {S.~J.}\ \bibnamefont {DeCamp}}, \bibinfo {author}
  {\bibfnamefont {L.}~\bibnamefont {Giomi}}, \bibinfo {author} {\bibfnamefont
  {M.~J.}\ \bibnamefont {Bowick}}, \bibinfo {author} {\bibfnamefont {M.~C.}\
  \bibnamefont {Marchetti}}, \bibinfo {author} {\bibfnamefont {Z.}~\bibnamefont
  {Dogic}},\ and\ \bibinfo {author} {\bibfnamefont {A.~R.}\ \bibnamefont
  {Bausch}},\ }\href {https://doi.org/10.1126/science.1254784} {\bibfield
  {journal} {\bibinfo  {journal} {Science}\ }\textbf {\bibinfo {volume}
  {345}},\ \bibinfo {pages} {1135} (\bibinfo {year} {2014})}\BibitemShut
  {NoStop}%
\bibitem [{\citenamefont {Metselaar}\ \emph {et~al.}(2019)\citenamefont
  {Metselaar}, \citenamefont {Yeomans},\ and\ \citenamefont
  {Doostmohammadi}}]{metselaar2019}%
  \BibitemOpen
  \bibfield  {author} {\bibinfo {author} {\bibfnamefont {L.}~\bibnamefont
  {Metselaar}}, \bibinfo {author} {\bibfnamefont {J.~M.}\ \bibnamefont
  {Yeomans}},\ and\ \bibinfo {author} {\bibfnamefont {A.}~\bibnamefont
  {Doostmohammadi}},\ }\href {https://doi.org/10.1103/physrevlett.123.208001}
  {\bibfield  {journal} {\bibinfo  {journal} {Phys. Rev. Lett.}\ }\textbf
  {\bibinfo {volume} {123}},\ \bibinfo {pages} {208001} (\bibinfo {year}
  {2019})}\BibitemShut {NoStop}%
\bibitem [{\citenamefont {Firouznia}\ and\ \citenamefont
  {Saintillan}(2025)}]{firouznia2025self}%
  \BibitemOpen
  \bibfield  {author} {\bibinfo {author} {\bibfnamefont {M.}~\bibnamefont
  {Firouznia}}\ and\ \bibinfo {author} {\bibfnamefont {D.}~\bibnamefont
  {Saintillan}},\ }\href {https://doi.org/10.1103/PhysRevResearch.7.L012054}
  {\bibfield  {journal} {\bibinfo  {journal} {Physical Review Research}\
  }\textbf {\bibinfo {volume} {7}},\ \bibinfo {pages} {L012054} (\bibinfo
  {year} {2025})}\BibitemShut {NoStop}%
\bibitem [{\citenamefont {Loiseau}\ \emph {et~al.}(2016)\citenamefont
  {Loiseau}, \citenamefont {Schneider}, \citenamefont {Keber}, \citenamefont
  {Pelzl}, \citenamefont {Massiera}, \citenamefont {Salbreux},\ and\
  \citenamefont {Bausch}}]{loiseau2016}%
  \BibitemOpen
  \bibfield  {author} {\bibinfo {author} {\bibfnamefont {E.}~\bibnamefont
  {Loiseau}}, \bibinfo {author} {\bibfnamefont {J.~A.~M.}\ \bibnamefont
  {Schneider}}, \bibinfo {author} {\bibfnamefont {F.~C.}\ \bibnamefont
  {Keber}}, \bibinfo {author} {\bibfnamefont {C.}~\bibnamefont {Pelzl}},
  \bibinfo {author} {\bibfnamefont {G.}~\bibnamefont {Massiera}}, \bibinfo
  {author} {\bibfnamefont {G.}~\bibnamefont {Salbreux}},\ and\ \bibinfo
  {author} {\bibfnamefont {A.~R.}\ \bibnamefont {Bausch}},\ }\href
  {https://doi.org/10.1126/sciadv.1500465} {\bibfield  {journal} {\bibinfo
  {journal} {Sci. Adv.}\ }\textbf {\bibinfo {volume} {2}},\ \bibinfo {pages}
  {e1500465} (\bibinfo {year} {2016})}\BibitemShut {NoStop}%
\bibitem [{\citenamefont {Litschel}\ \emph {et~al.}(2021)\citenamefont
  {Litschel}, \citenamefont {Kelley}, \citenamefont {Holz}, \citenamefont
  {Adeli~Koudehi}, \citenamefont {Vogel}, \citenamefont {Burbaum},
  \citenamefont {Mizuno}, \citenamefont {Vavylonis},\ and\ \citenamefont
  {Schwille}}]{litschel2021}%
  \BibitemOpen
  \bibfield  {author} {\bibinfo {author} {\bibfnamefont {T.}~\bibnamefont
  {Litschel}}, \bibinfo {author} {\bibfnamefont {C.~F.}\ \bibnamefont
  {Kelley}}, \bibinfo {author} {\bibfnamefont {D.}~\bibnamefont {Holz}},
  \bibinfo {author} {\bibfnamefont {M.}~\bibnamefont {Adeli~Koudehi}}, \bibinfo
  {author} {\bibfnamefont {S.~K.}\ \bibnamefont {Vogel}}, \bibinfo {author}
  {\bibfnamefont {L.}~\bibnamefont {Burbaum}}, \bibinfo {author} {\bibfnamefont
  {N.}~\bibnamefont {Mizuno}}, \bibinfo {author} {\bibfnamefont
  {D.}~\bibnamefont {Vavylonis}},\ and\ \bibinfo {author} {\bibfnamefont
  {P.}~\bibnamefont {Schwille}},\ }\href
  {https://doi.org/10.1038/s41467-021-22422-7} {\bibfield  {journal} {\bibinfo
  {journal} {Nat. Commun.}\ }\textbf {\bibinfo {volume} {12}},\ \bibinfo
  {pages} {2254} (\bibinfo {year} {2021})}\BibitemShut {NoStop}%
\bibitem [{\citenamefont {Takatori}\ and\ \citenamefont
  {Sahu}(2020)}]{takatori2020active}%
  \BibitemOpen
  \bibfield  {author} {\bibinfo {author} {\bibfnamefont {S.~C.}\ \bibnamefont
  {Takatori}}\ and\ \bibinfo {author} {\bibfnamefont {A.}~\bibnamefont
  {Sahu}},\ }\href {https://doi.org/10.1103/PhysRevLett.124.158102} {\bibfield
  {journal} {\bibinfo  {journal} {Physical review letters}\ }\textbf {\bibinfo
  {volume} {124}},\ \bibinfo {pages} {158102} (\bibinfo {year}
  {2020})}\BibitemShut {NoStop}%
\bibitem [{\citenamefont {Vutukuri}\ \emph {et~al.}(2020)\citenamefont
  {Vutukuri}, \citenamefont {Hoore}, \citenamefont {Abaurrea-Velasco},
  \citenamefont {van Buren}, \citenamefont {Dutto}, \citenamefont {Auth},
  \citenamefont {Fedosov}, \citenamefont {Gompper},\ and\ \citenamefont
  {Vermant}}]{vutukuri2020}%
  \BibitemOpen
  \bibfield  {author} {\bibinfo {author} {\bibfnamefont {H.~R.}\ \bibnamefont
  {Vutukuri}}, \bibinfo {author} {\bibfnamefont {M.}~\bibnamefont {Hoore}},
  \bibinfo {author} {\bibfnamefont {C.}~\bibnamefont {Abaurrea-Velasco}},
  \bibinfo {author} {\bibfnamefont {L.}~\bibnamefont {van Buren}}, \bibinfo
  {author} {\bibfnamefont {A.}~\bibnamefont {Dutto}}, \bibinfo {author}
  {\bibfnamefont {T.}~\bibnamefont {Auth}}, \bibinfo {author} {\bibfnamefont
  {D.~A.}\ \bibnamefont {Fedosov}}, \bibinfo {author} {\bibfnamefont
  {G.}~\bibnamefont {Gompper}},\ and\ \bibinfo {author} {\bibfnamefont
  {J.}~\bibnamefont {Vermant}},\ }\href
  {https://doi.org/10.1038/s41586-020-2730-x} {\bibfield  {journal} {\bibinfo
  {journal} {Nature}\ }\textbf {\bibinfo {volume} {586}},\ \bibinfo {pages}
  {52} (\bibinfo {year} {2020})}\BibitemShut {NoStop}%
\bibitem [{\citenamefont {Le~Nagard}\ \emph {et~al.}(2022)\citenamefont
  {Le~Nagard}, \citenamefont {Brown}, \citenamefont {Dawson}, \citenamefont
  {Martinez}, \citenamefont {Poon},\ and\ \citenamefont
  {Staykova}}]{le2022encapsulated}%
  \BibitemOpen
  \bibfield  {author} {\bibinfo {author} {\bibfnamefont {L.}~\bibnamefont
  {Le~Nagard}}, \bibinfo {author} {\bibfnamefont {A.~T.}\ \bibnamefont
  {Brown}}, \bibinfo {author} {\bibfnamefont {A.}~\bibnamefont {Dawson}},
  \bibinfo {author} {\bibfnamefont {V.~A.}\ \bibnamefont {Martinez}}, \bibinfo
  {author} {\bibfnamefont {W.~C.}\ \bibnamefont {Poon}},\ and\ \bibinfo
  {author} {\bibfnamefont {M.}~\bibnamefont {Staykova}},\ }\href
  {https://doi.org/10.1073/pnas.2206096119} {\bibfield  {journal} {\bibinfo
  {journal} {Proceedings of the National Academy of Sciences}\ }\textbf
  {\bibinfo {volume} {119}},\ \bibinfo {pages} {e2206096119} (\bibinfo {year}
  {2022})}\BibitemShut {NoStop}%
\bibitem [{\citenamefont {Paoluzzi}\ \emph {et~al.}(2016)\citenamefont
  {Paoluzzi}, \citenamefont {Di~Leonardo}, \citenamefont {Marchetti},\ and\
  \citenamefont {Angelani}}]{paoluzzi2016}%
  \BibitemOpen
  \bibfield  {author} {\bibinfo {author} {\bibfnamefont {M.}~\bibnamefont
  {Paoluzzi}}, \bibinfo {author} {\bibfnamefont {R.}~\bibnamefont
  {Di~Leonardo}}, \bibinfo {author} {\bibfnamefont {M.~C.}\ \bibnamefont
  {Marchetti}},\ and\ \bibinfo {author} {\bibfnamefont {L.}~\bibnamefont
  {Angelani}},\ }\href {https://doi.org/10.1038/srep34146} {\bibfield
  {journal} {\bibinfo  {journal} {Sci. Rep.}\ }\textbf {\bibinfo {volume}
  {6}},\ \bibinfo {pages} {34146} (\bibinfo {year} {2016})}\BibitemShut
  {NoStop}%
\bibitem [{\citenamefont {Peterson}\ \emph {et~al.}(2021)\citenamefont
  {Peterson}, \citenamefont {Baskaran},\ and\ \citenamefont
  {Hagan}}]{peterson2021vesicle}%
  \BibitemOpen
  \bibfield  {author} {\bibinfo {author} {\bibfnamefont {M.~S.}\ \bibnamefont
  {Peterson}}, \bibinfo {author} {\bibfnamefont {A.}~\bibnamefont {Baskaran}},\
  and\ \bibinfo {author} {\bibfnamefont {M.~F.}\ \bibnamefont {Hagan}},\ }\href
  {https://doi.org/10.1038/s41467-021-27310-8} {\bibfield  {journal} {\bibinfo
  {journal} {Nature communications}\ }\textbf {\bibinfo {volume} {12}},\
  \bibinfo {pages} {7247} (\bibinfo {year} {2021})}\BibitemShut {NoStop}%
\bibitem [{\citenamefont {Wioland}\ \emph {et~al.}(2016)\citenamefont
  {Wioland}, \citenamefont {Woodhouse}, \citenamefont {Dunkel},\ and\
  \citenamefont {Goldstein}}]{wioland2016ferromagnetic}%
  \BibitemOpen
  \bibfield  {author} {\bibinfo {author} {\bibfnamefont {H.}~\bibnamefont
  {Wioland}}, \bibinfo {author} {\bibfnamefont {F.~G.}\ \bibnamefont
  {Woodhouse}}, \bibinfo {author} {\bibfnamefont {J.}~\bibnamefont {Dunkel}},\
  and\ \bibinfo {author} {\bibfnamefont {R.~E.}\ \bibnamefont {Goldstein}},\
  }\href {https://doi.org/10.1038/nphys3607} {\bibfield  {journal} {\bibinfo
  {journal} {Nature physics}\ }\textbf {\bibinfo {volume} {12}},\ \bibinfo
  {pages} {341} (\bibinfo {year} {2016})}\BibitemShut {NoStop}%
\bibitem [{\citenamefont {Doostmohammadi}\ \emph {et~al.}(2017)\citenamefont
  {Doostmohammadi}, \citenamefont {Shendruk}, \citenamefont {Thijssen},\ and\
  \citenamefont {Yeomans}}]{doostmohammadi2017onset}%
  \BibitemOpen
  \bibfield  {author} {\bibinfo {author} {\bibfnamefont {A.}~\bibnamefont
  {Doostmohammadi}}, \bibinfo {author} {\bibfnamefont {T.~N.}\ \bibnamefont
  {Shendruk}}, \bibinfo {author} {\bibfnamefont {K.}~\bibnamefont {Thijssen}},\
  and\ \bibinfo {author} {\bibfnamefont {J.~M.}\ \bibnamefont {Yeomans}},\
  }\href {https://doi.org/10.1038/ncomms15326} {\bibfield  {journal} {\bibinfo
  {journal} {Nature communications}\ }\textbf {\bibinfo {volume} {8}},\
  \bibinfo {pages} {15326} (\bibinfo {year} {2017})}\BibitemShut {NoStop}%
\bibitem [{\citenamefont {Wu}\ \emph {et~al.}(2017)\citenamefont {Wu},
  \citenamefont {Hishamunda}, \citenamefont {Chen}, \citenamefont {DeCamp},
  \citenamefont {Chang}, \citenamefont {Fern{\'a}ndez-Nieves}, \citenamefont
  {Fraden},\ and\ \citenamefont {Dogic}}]{wu2017}%
  \BibitemOpen
  \bibfield  {author} {\bibinfo {author} {\bibfnamefont {K.-T.}\ \bibnamefont
  {Wu}}, \bibinfo {author} {\bibfnamefont {J.~B.}\ \bibnamefont {Hishamunda}},
  \bibinfo {author} {\bibfnamefont {D.~T.~N.}\ \bibnamefont {Chen}}, \bibinfo
  {author} {\bibfnamefont {S.~J.}\ \bibnamefont {DeCamp}}, \bibinfo {author}
  {\bibfnamefont {Y.-W.}\ \bibnamefont {Chang}}, \bibinfo {author}
  {\bibfnamefont {A.}~\bibnamefont {Fern{\'a}ndez-Nieves}}, \bibinfo {author}
  {\bibfnamefont {S.}~\bibnamefont {Fraden}},\ and\ \bibinfo {author}
  {\bibfnamefont {Z.}~\bibnamefont {Dogic}},\ }\href
  {https://doi.org/10.1126/science.aal1979} {\bibfield  {journal} {\bibinfo
  {journal} {Science}\ }\textbf {\bibinfo {volume} {355}},\ \bibinfo {pages}
  {eaal1979} (\bibinfo {year} {2017})}\BibitemShut {NoStop}%
\bibitem [{\citenamefont {Theillard}\ \emph {et~al.}(2017)\citenamefont
  {Theillard}, \citenamefont {Alonso-Matilla},\ and\ \citenamefont
  {Saintillan}}]{theillard2017geometric}%
  \BibitemOpen
  \bibfield  {author} {\bibinfo {author} {\bibfnamefont {M.}~\bibnamefont
  {Theillard}}, \bibinfo {author} {\bibfnamefont {R.}~\bibnamefont
  {Alonso-Matilla}},\ and\ \bibinfo {author} {\bibfnamefont {D.}~\bibnamefont
  {Saintillan}},\ }\href {https://doi.org/10.1039/c6sm01955b} {\bibfield
  {journal} {\bibinfo  {journal} {Soft Matter}\ }\textbf {\bibinfo {volume}
  {13}},\ \bibinfo {pages} {363} (\bibinfo {year} {2017})}\BibitemShut
  {NoStop}%
\bibitem [{\citenamefont {Shendruk}\ \emph {et~al.}(2017)\citenamefont
  {Shendruk}, \citenamefont {Doostmohammadi}, \citenamefont {Thijssen},\ and\
  \citenamefont {Yeomans}}]{shendruk2017}%
  \BibitemOpen
  \bibfield  {author} {\bibinfo {author} {\bibfnamefont {T.~N.}\ \bibnamefont
  {Shendruk}}, \bibinfo {author} {\bibfnamefont {A.}~\bibnamefont
  {Doostmohammadi}}, \bibinfo {author} {\bibfnamefont {K.}~\bibnamefont
  {Thijssen}},\ and\ \bibinfo {author} {\bibfnamefont {J.~M.}\ \bibnamefont
  {Yeomans}},\ }\href {https://doi.org/10.1039/c6sm02310j} {\bibfield
  {journal} {\bibinfo  {journal} {Soft Matter}\ }\textbf {\bibinfo {volume}
  {13}},\ \bibinfo {pages} {3853} (\bibinfo {year} {2017})}\BibitemShut
  {NoStop}%
\bibitem [{\citenamefont {Norton}\ \emph {et~al.}(2018)\citenamefont {Norton},
  \citenamefont {Baskaran}, \citenamefont {Opathalage}, \citenamefont
  {Langeslay}, \citenamefont {Fraden}, \citenamefont {Baskaran},\ and\
  \citenamefont {Hagan}}]{norton2018}%
  \BibitemOpen
  \bibfield  {author} {\bibinfo {author} {\bibfnamefont {M.~M.}\ \bibnamefont
  {Norton}}, \bibinfo {author} {\bibfnamefont {A.}~\bibnamefont {Baskaran}},
  \bibinfo {author} {\bibfnamefont {A.}~\bibnamefont {Opathalage}}, \bibinfo
  {author} {\bibfnamefont {B.}~\bibnamefont {Langeslay}}, \bibinfo {author}
  {\bibfnamefont {S.}~\bibnamefont {Fraden}}, \bibinfo {author} {\bibfnamefont
  {A.}~\bibnamefont {Baskaran}},\ and\ \bibinfo {author} {\bibfnamefont
  {M.~F.}\ \bibnamefont {Hagan}},\ }\href
  {https://doi.org/10.1103/physreve.97.012702} {\bibfield  {journal} {\bibinfo
  {journal} {Phys. Rev. E}\ }\textbf {\bibinfo {volume} {97}},\ \bibinfo
  {pages} {012702} (\bibinfo {year} {2018})}\BibitemShut {NoStop}%
\bibitem [{\citenamefont {Opathalage}\ \emph {et~al.}(2019)\citenamefont
  {Opathalage}, \citenamefont {Norton}, \citenamefont {Juniper}, \citenamefont
  {Langeslay}, \citenamefont {Aghvami}, \citenamefont {Fraden},\ and\
  \citenamefont {Dogic}}]{opathalage2019}%
  \BibitemOpen
  \bibfield  {author} {\bibinfo {author} {\bibfnamefont {A.}~\bibnamefont
  {Opathalage}}, \bibinfo {author} {\bibfnamefont {M.~M.}\ \bibnamefont
  {Norton}}, \bibinfo {author} {\bibfnamefont {M.~P.~N.}\ \bibnamefont
  {Juniper}}, \bibinfo {author} {\bibfnamefont {B.}~\bibnamefont {Langeslay}},
  \bibinfo {author} {\bibfnamefont {S.~A.}\ \bibnamefont {Aghvami}}, \bibinfo
  {author} {\bibfnamefont {S.}~\bibnamefont {Fraden}},\ and\ \bibinfo {author}
  {\bibfnamefont {Z.}~\bibnamefont {Dogic}},\ }\href
  {https://doi.org/10.1073/pnas.1816733116} {\bibfield  {journal} {\bibinfo
  {journal} {Proc. Natl. Acad. Sci. U.S.A.}\ }\textbf {\bibinfo {volume}
  {116}},\ \bibinfo {pages} {4788} (\bibinfo {year} {2019})}\BibitemShut
  {NoStop}%
\bibitem [{\citenamefont {Hardo{\"u}in}\ \emph {et~al.}(2019)\citenamefont
  {Hardo{\"u}in}, \citenamefont {Hughes}, \citenamefont {Doostmohammadi},
  \citenamefont {Laurent}, \citenamefont {Lopez-Leon}, \citenamefont {Yeomans},
  \citenamefont {Ign{\'e}s-Mullol},\ and\ \citenamefont
  {Sagu{\'e}s}}]{hardouin2019}%
  \BibitemOpen
  \bibfield  {author} {\bibinfo {author} {\bibfnamefont {J.}~\bibnamefont
  {Hardo{\"u}in}}, \bibinfo {author} {\bibfnamefont {R.}~\bibnamefont
  {Hughes}}, \bibinfo {author} {\bibfnamefont {A.}~\bibnamefont
  {Doostmohammadi}}, \bibinfo {author} {\bibfnamefont {J.}~\bibnamefont
  {Laurent}}, \bibinfo {author} {\bibfnamefont {T.}~\bibnamefont {Lopez-Leon}},
  \bibinfo {author} {\bibfnamefont {J.~M.}\ \bibnamefont {Yeomans}}, \bibinfo
  {author} {\bibfnamefont {J.}~\bibnamefont {Ign{\'e}s-Mullol}},\ and\ \bibinfo
  {author} {\bibfnamefont {F.}~\bibnamefont {Sagu{\'e}s}},\ }\href
  {https://doi.org/10.1038/s42005-019-0221-x} {\bibfield  {journal} {\bibinfo
  {journal} {Commun. Phys.}\ }\textbf {\bibinfo {volume} {2}},\ \bibinfo
  {pages} {121} (\bibinfo {year} {2019})}\BibitemShut {NoStop}%
\bibitem [{\citenamefont {Alert}(2022)}]{alert2022fingering}%
  \BibitemOpen
  \bibfield  {author} {\bibinfo {author} {\bibfnamefont {R.}~\bibnamefont
  {Alert}},\ }\href {https://doi.org/10.1088/1751-8121/ac6c61} {\bibfield
  {journal} {\bibinfo  {journal} {Journal of Physics A: Mathematical and
  Theoretical}\ }\textbf {\bibinfo {volume} {55}},\ \bibinfo {pages} {234009}
  (\bibinfo {year} {2022})}\BibitemShut {NoStop}%
\bibitem [{\citenamefont {Dhar}\ \emph {et~al.}(2025)\citenamefont {Dhar},
  \citenamefont {Shelley},\ and\ \citenamefont {Saintillan}}]{dhar2025}%
  \BibitemOpen
  \bibfield  {author} {\bibinfo {author} {\bibfnamefont {T.}~\bibnamefont
  {Dhar}}, \bibinfo {author} {\bibfnamefont {M.~J.}\ \bibnamefont {Shelley}},\
  and\ \bibinfo {author} {\bibfnamefont {D.}~\bibnamefont {Saintillan}},\
  }\href {https://doi.org/10.1103/m5ck-v2q3} {\bibfield  {journal} {\bibinfo
  {journal} {Phys. Rev. Fluids}\ }\textbf {\bibinfo {volume} {10}},\ \bibinfo
  {pages} {083103} (\bibinfo {year} {2025})}\BibitemShut {NoStop}%
\bibitem [{\citenamefont {Chandler}\ and\ \citenamefont
  {Spagnolie}(2024)}]{chandler2024}%
  \BibitemOpen
  \bibfield  {author} {\bibinfo {author} {\bibfnamefont {T.~G.~J.}\
  \bibnamefont {Chandler}}\ and\ \bibinfo {author} {\bibfnamefont {S.~E.}\
  \bibnamefont {Spagnolie}},\ }\href
  {https://doi.org/10.1103/physrevfluids.9.110511} {\bibfield  {journal}
  {\bibinfo  {journal} {Phys. Rev. Fluids}\ }\textbf {\bibinfo {volume} {9}},\
  \bibinfo {pages} {110511} (\bibinfo {year} {2024})}\BibitemShut {NoStop}%
\bibitem [{\citenamefont {Chandler}\ and\ \citenamefont
  {Spagnolie}(2025)}]{chandler2025}%
  \BibitemOpen
  \bibfield  {author} {\bibinfo {author} {\bibfnamefont {T.~G.~J.}\
  \bibnamefont {Chandler}}\ and\ \bibinfo {author} {\bibfnamefont {S.~E.}\
  \bibnamefont {Spagnolie}},\ }\href {https://doi.org/10.1103/x8y2-tw4c}
  {\bibfield  {journal} {\bibinfo  {journal} {Phys. Rev. Res.}\ }\textbf
  {\bibinfo {volume} {7}},\ \bibinfo {pages} {023263} (\bibinfo {year}
  {2025})}\BibitemShut {NoStop}%
\bibitem [{\citenamefont {Saintillan}\ and\ \citenamefont
  {Shelley}(2008)}]{saintillan2008prl}%
  \BibitemOpen
  \bibfield  {author} {\bibinfo {author} {\bibfnamefont {D.}~\bibnamefont
  {Saintillan}}\ and\ \bibinfo {author} {\bibfnamefont {M.~J.}\ \bibnamefont
  {Shelley}},\ }\href {https://doi.org/10.1103/physrevlett.100.178103}
  {\bibfield  {journal} {\bibinfo  {journal} {Phys. Rev. Lett.}\ }\textbf
  {\bibinfo {volume} {100}},\ \bibinfo {pages} {178103} (\bibinfo {year}
  {2008})}\BibitemShut {NoStop}%
\bibitem [{\citenamefont {Peskin}(2002)}]{peskin2002}%
  \BibitemOpen
  \bibfield  {author} {\bibinfo {author} {\bibfnamefont {C.~S.}\ \bibnamefont
  {Peskin}},\ }\href {https://doi.org/10.1017/s0962492902000077} {\bibfield
  {journal} {\bibinfo  {journal} {Acta Numer.}\ }\textbf {\bibinfo {volume}
  {11}},\ \bibinfo {pages} {479} (\bibinfo {year} {2002})}\BibitemShut
  {NoStop}%
\bibitem [{\citenamefont {Peskin}\ and\ \citenamefont
  {Printz}(1993)}]{peskinprintz1993}%
  \BibitemOpen
  \bibfield  {author} {\bibinfo {author} {\bibfnamefont {C.~S.}\ \bibnamefont
  {Peskin}}\ and\ \bibinfo {author} {\bibfnamefont {B.~F.}\ \bibnamefont
  {Printz}},\ }\href {https://doi.org/10.1006/jcph.1993.1051} {\bibfield
  {journal} {\bibinfo  {journal} {J. Comput. Phys.}\ }\textbf {\bibinfo
  {volume} {105}},\ \bibinfo {pages} {33} (\bibinfo {year} {1993})}\BibitemShut
  {NoStop}%
\bibitem [{\citenamefont {Gennes}\ and\ \citenamefont
  {Prost}(1993)}]{de1995physics}%
  \BibitemOpen
  \bibfield  {author} {\bibinfo {author} {\bibfnamefont {P.~G.~D.}\
  \bibnamefont {Gennes}}\ and\ \bibinfo {author} {\bibfnamefont
  {J.}~\bibnamefont {Prost}},\ }\href
  {https://doi.org/10.1093/oso/9780198520245.001.0001} {\emph {\bibinfo {title}
  {The Physics of Liquid Crystals}}}\ (\bibinfo  {publisher} {Oxford University
  Press},\ \bibinfo {year} {1993})\BibitemShut {NoStop}%
\bibitem [{\citenamefont {Chandler}\ and\ \citenamefont
  {Spagnolie}(2023)}]{chandler2023}%
  \BibitemOpen
  \bibfield  {author} {\bibinfo {author} {\bibfnamefont {T.~G.}\ \bibnamefont
  {Chandler}}\ and\ \bibinfo {author} {\bibfnamefont {S.~E.}\ \bibnamefont
  {Spagnolie}},\ }\href {https://doi.org/10.1017/jfm.2023.488} {\bibfield
  {journal} {\bibinfo  {journal} {J. Fluid Mech.}\ }\textbf {\bibinfo {volume}
  {967}},\ \bibinfo {pages} {A19} (\bibinfo {year} {2023})}\BibitemShut
  {NoStop}%
\bibitem [{\citenamefont {Kantsler}\ and\ \citenamefont
  {Steinberg}(2005)}]{kantsler2005}%
  \BibitemOpen
  \bibfield  {author} {\bibinfo {author} {\bibfnamefont {V.}~\bibnamefont
  {Kantsler}}\ and\ \bibinfo {author} {\bibfnamefont {V.}~\bibnamefont
  {Steinberg}},\ }\href {https://doi.org/10.1103/physrevlett.95.258101}
  {\bibfield  {journal} {\bibinfo  {journal} {Phys. Rev. Lett.}\ }\textbf
  {\bibinfo {volume} {95}},\ \bibinfo {pages} {258101} (\bibinfo {year}
  {2005})}\BibitemShut {NoStop}%
\bibitem [{\citenamefont {Liu}\ \emph {et~al.}(2018)\citenamefont {Liu},
  \citenamefont {Chakrabarti}, \citenamefont {Saintillan}, \citenamefont
  {Lindner},\ and\ \citenamefont {du~Roure}}]{liu2018morphological}%
  \BibitemOpen
  \bibfield  {author} {\bibinfo {author} {\bibfnamefont {Y.}~\bibnamefont
  {Liu}}, \bibinfo {author} {\bibfnamefont {B.}~\bibnamefont {Chakrabarti}},
  \bibinfo {author} {\bibfnamefont {D.}~\bibnamefont {Saintillan}}, \bibinfo
  {author} {\bibfnamefont {A.}~\bibnamefont {Lindner}},\ and\ \bibinfo {author}
  {\bibfnamefont {O.}~\bibnamefont {du~Roure}},\ }\href
  {https://doi.org/10.1073/pnas.1805399115} {\bibfield  {journal} {\bibinfo
  {journal} {Proceedings of the National Academy of Sciences}\ }\textbf
  {\bibinfo {volume} {115}},\ \bibinfo {pages} {9438} (\bibinfo {year}
  {2018})}\BibitemShut {NoStop}%
\bibitem [{\citenamefont {Chakrabarti}\ \emph {et~al.}(2023)\citenamefont
  {Chakrabarti}, \citenamefont {Shelley},\ and\ \citenamefont
  {F{\"u}rthauer}}]{chakrabarti2023collective}%
  \BibitemOpen
  \bibfield  {author} {\bibinfo {author} {\bibfnamefont {B.}~\bibnamefont
  {Chakrabarti}}, \bibinfo {author} {\bibfnamefont {M.~J.}\ \bibnamefont
  {Shelley}},\ and\ \bibinfo {author} {\bibfnamefont {S.}~\bibnamefont
  {F{\"u}rthauer}},\ }\href {https://doi.org/10.1103/PhysRevLett.130.128202}
  {\bibfield  {journal} {\bibinfo  {journal} {Phys. Rev. Lett.}\ }\textbf
  {\bibinfo {volume} {130}},\ \bibinfo {pages} {128202} (\bibinfo {year}
  {2023})}\BibitemShut {NoStop}%
\bibitem [{\citenamefont {Kim}\ \emph {et~al.}(2021)\citenamefont {Kim},
  \citenamefont {Pochitaloff}, \citenamefont {Stooke-Vaughan},\ and\
  \citenamefont {Camp{\`a}s}}]{kim2021}%
  \BibitemOpen
  \bibfield  {author} {\bibinfo {author} {\bibfnamefont {S.}~\bibnamefont
  {Kim}}, \bibinfo {author} {\bibfnamefont {M.}~\bibnamefont {Pochitaloff}},
  \bibinfo {author} {\bibfnamefont {G.~A.}\ \bibnamefont {Stooke-Vaughan}},\
  and\ \bibinfo {author} {\bibfnamefont {O.}~\bibnamefont {Camp{\`a}s}},\
  }\href {https://doi.org/10.1038/s41567-021-01215-1} {\bibfield  {journal}
  {\bibinfo  {journal} {Nat. Phys.}\ }\textbf {\bibinfo {volume} {17}},\
  \bibinfo {pages} {859} (\bibinfo {year} {2021})}\BibitemShut {NoStop}%
\bibitem [{\citenamefont {Bi}\ \emph {et~al.}(2016)\citenamefont {Bi},
  \citenamefont {Yang}, \citenamefont {Marchetti},\ and\ \citenamefont
  {Manning}}]{bi2016}%
  \BibitemOpen
  \bibfield  {author} {\bibinfo {author} {\bibfnamefont {D.}~\bibnamefont
  {Bi}}, \bibinfo {author} {\bibfnamefont {X.}~\bibnamefont {Yang}}, \bibinfo
  {author} {\bibfnamefont {M.~C.}\ \bibnamefont {Marchetti}},\ and\ \bibinfo
  {author} {\bibfnamefont {M.~L.}\ \bibnamefont {Manning}},\ }\href
  {https://doi.org/10.1103/physrevx.6.021011} {\bibfield  {journal} {\bibinfo
  {journal} {Phys. Rev. X}\ }\textbf {\bibinfo {volume} {6}},\ \bibinfo {pages}
  {021011} (\bibinfo {year} {2016})}\BibitemShut {NoStop}%
\bibitem [{\citenamefont {Prost}\ \emph {et~al.}(2015)\citenamefont {Prost},
  \citenamefont {J{\"u}licher},\ and\ \citenamefont {Joanny}}]{prost2015}%
  \BibitemOpen
  \bibfield  {author} {\bibinfo {author} {\bibfnamefont {J.}~\bibnamefont
  {Prost}}, \bibinfo {author} {\bibfnamefont {F.}~\bibnamefont
  {J{\"u}licher}},\ and\ \bibinfo {author} {\bibfnamefont {J.-F.}\ \bibnamefont
  {Joanny}},\ }\href {https://doi.org/10.1038/nphys3224} {\bibfield  {journal}
  {\bibinfo  {journal} {Nat. Phys.}\ }\textbf {\bibinfo {volume} {11}},\
  \bibinfo {pages} {111} (\bibinfo {year} {2015})}\BibitemShut {NoStop}%
\bibitem [{\citenamefont {Villalobos-Concha}\ \emph {et~al.}(2025)\citenamefont
  {Villalobos-Concha}, \citenamefont {Liu}, \citenamefont {Ramos},
  \citenamefont {Goral}, \citenamefont {Lindner}, \citenamefont {López-León},
  \citenamefont {Clément}, \citenamefont {Soto},\ and\ \citenamefont
  {Cordero}}]{villalobos2025active}%
  \BibitemOpen
  \bibfield  {author} {\bibinfo {author} {\bibfnamefont {C.}~\bibnamefont
  {Villalobos-Concha}}, \bibinfo {author} {\bibfnamefont {Z.}~\bibnamefont
  {Liu}}, \bibinfo {author} {\bibfnamefont {G.}~\bibnamefont {Ramos}}, \bibinfo
  {author} {\bibfnamefont {M.}~\bibnamefont {Goral}}, \bibinfo {author}
  {\bibfnamefont {A.}~\bibnamefont {Lindner}}, \bibinfo {author} {\bibfnamefont
  {T.}~\bibnamefont {López-León}}, \bibinfo {author} {\bibfnamefont
  {E.}~\bibnamefont {Clément}}, \bibinfo {author} {\bibfnamefont
  {R.}~\bibnamefont {Soto}},\ and\ \bibinfo {author} {\bibfnamefont {M.~L.}\
  \bibnamefont {Cordero}},\ }\href {https://doi.org/10.1073/pnas.2426096122}
  {\bibfield  {journal} {\bibinfo  {journal} {Proceedings of the National
  Academy of Sciences}\ }\textbf {\bibinfo {volume} {122}},\ \bibinfo {pages}
  {e2426096122} (\bibinfo {year} {2025})}\BibitemShut {NoStop}%
\bibitem [{\citenamefont {Brochard}\ and\ \citenamefont
  {Lennon}(1975)}]{brochard1975}%
  \BibitemOpen
  \bibfield  {author} {\bibinfo {author} {\bibfnamefont {F.}~\bibnamefont
  {Brochard}}\ and\ \bibinfo {author} {\bibfnamefont {J.}~\bibnamefont
  {Lennon}},\ }\href {https://doi.org/10.1051/jphys:0197500360110103500}
  {\bibfield  {journal} {\bibinfo  {journal} {J. Phys. (Paris)}\ }\textbf
  {\bibinfo {volume} {36}},\ \bibinfo {pages} {1035} (\bibinfo {year}
  {1975})}\BibitemShut {NoStop}%
\bibitem [{\citenamefont {Betz}\ \emph {et~al.}(2009)\citenamefont {Betz},
  \citenamefont {Lenz}, \citenamefont {Joanny},\ and\ \citenamefont
  {Sykes}}]{betz2009}%
  \BibitemOpen
  \bibfield  {author} {\bibinfo {author} {\bibfnamefont {T.}~\bibnamefont
  {Betz}}, \bibinfo {author} {\bibfnamefont {M.}~\bibnamefont {Lenz}}, \bibinfo
  {author} {\bibfnamefont {J.-F.}\ \bibnamefont {Joanny}},\ and\ \bibinfo
  {author} {\bibfnamefont {C.}~\bibnamefont {Sykes}},\ }\href
  {https://doi.org/10.1073/pnas.0904614106} {\bibfield  {journal} {\bibinfo
  {journal} {Proc. Natl. Acad. Sci. U.S.A.}\ }\textbf {\bibinfo {volume}
  {106}},\ \bibinfo {pages} {15320} (\bibinfo {year} {2009})}\BibitemShut
  {NoStop}%
\bibitem [{\citenamefont {Turlier}\ \emph {et~al.}(2016)\citenamefont
  {Turlier}, \citenamefont {Fedosov}, \citenamefont {Audoly}, \citenamefont
  {Auth}, \citenamefont {Gov}, \citenamefont {Sykes}, \citenamefont {Joanny},
  \citenamefont {Gompper},\ and\ \citenamefont {Betz}}]{turlier2016}%
  \BibitemOpen
  \bibfield  {author} {\bibinfo {author} {\bibfnamefont {H.}~\bibnamefont
  {Turlier}}, \bibinfo {author} {\bibfnamefont {D.~A.}\ \bibnamefont
  {Fedosov}}, \bibinfo {author} {\bibfnamefont {B.}~\bibnamefont {Audoly}},
  \bibinfo {author} {\bibfnamefont {T.}~\bibnamefont {Auth}}, \bibinfo {author}
  {\bibfnamefont {N.~S.}\ \bibnamefont {Gov}}, \bibinfo {author} {\bibfnamefont
  {C.}~\bibnamefont {Sykes}}, \bibinfo {author} {\bibfnamefont {J.-F.}\
  \bibnamefont {Joanny}}, \bibinfo {author} {\bibfnamefont {G.}~\bibnamefont
  {Gompper}},\ and\ \bibinfo {author} {\bibfnamefont {T.}~\bibnamefont
  {Betz}},\ }\href {https://doi.org/10.1038/nphys3621} {\bibfield  {journal}
  {\bibinfo  {journal} {Nat. Phys.}\ }\textbf {\bibinfo {volume} {12}},\
  \bibinfo {pages} {513} (\bibinfo {year} {2016})}\BibitemShut {NoStop}%
\bibitem [{\citenamefont {Zhang}\ \emph {et~al.}(2021)\citenamefont {Zhang},
  \citenamefont {Redford}, \citenamefont {Ruijgrok}, \citenamefont {Kumar},
  \citenamefont {Mozaffari}, \citenamefont {Zemsky}, \citenamefont {Dinner},
  \citenamefont {Vitelli}, \citenamefont {Bryant}, \citenamefont {Gardel},\
  and\ \citenamefont {de~Pablo}}]{zhang2021}%
  \BibitemOpen
  \bibfield  {author} {\bibinfo {author} {\bibfnamefont {R.}~\bibnamefont
  {Zhang}}, \bibinfo {author} {\bibfnamefont {S.~A.}\ \bibnamefont {Redford}},
  \bibinfo {author} {\bibfnamefont {P.~V.}\ \bibnamefont {Ruijgrok}}, \bibinfo
  {author} {\bibfnamefont {N.}~\bibnamefont {Kumar}}, \bibinfo {author}
  {\bibfnamefont {A.}~\bibnamefont {Mozaffari}}, \bibinfo {author}
  {\bibfnamefont {S.}~\bibnamefont {Zemsky}}, \bibinfo {author} {\bibfnamefont
  {A.~R.}\ \bibnamefont {Dinner}}, \bibinfo {author} {\bibfnamefont
  {V.}~\bibnamefont {Vitelli}}, \bibinfo {author} {\bibfnamefont
  {Z.}~\bibnamefont {Bryant}}, \bibinfo {author} {\bibfnamefont {M.~L.}\
  \bibnamefont {Gardel}},\ and\ \bibinfo {author} {\bibfnamefont {J.~J.}\
  \bibnamefont {de~Pablo}},\ }\href
  {https://doi.org/10.1038/s41563-020-00901-4} {\bibfield  {journal} {\bibinfo
  {journal} {Nat. Mater.}\ }\textbf {\bibinfo {volume} {20}},\ \bibinfo {pages}
  {875} (\bibinfo {year} {2021})}\BibitemShut {NoStop}%
\bibitem [{\citenamefont {Lemma}\ \emph {et~al.}(2023)\citenamefont {Lemma},
  \citenamefont {Varghese}, \citenamefont {Ross}, \citenamefont {Thomson},
  \citenamefont {Baskaran},\ and\ \citenamefont {Dogic}}]{lemma2023}%
  \BibitemOpen
  \bibfield  {author} {\bibinfo {author} {\bibfnamefont {L.~M.}\ \bibnamefont
  {Lemma}}, \bibinfo {author} {\bibfnamefont {M.}~\bibnamefont {Varghese}},
  \bibinfo {author} {\bibfnamefont {T.~D.}\ \bibnamefont {Ross}}, \bibinfo
  {author} {\bibfnamefont {M.}~\bibnamefont {Thomson}}, \bibinfo {author}
  {\bibfnamefont {A.}~\bibnamefont {Baskaran}},\ and\ \bibinfo {author}
  {\bibfnamefont {Z.}~\bibnamefont {Dogic}},\ }\href
  {https://doi.org/10.1093/pnasnexus/pgad130} {\bibfield  {journal} {\bibinfo
  {journal} {PNAS Nexus}\ }\textbf {\bibinfo {volume} {2}},\ \bibinfo {pages}
  {pgad130} (\bibinfo {year} {2023})}\BibitemShut {NoStop}%
\end{thebibliography}%

\end{document}